\documentclass[aps,pra,twocolumn,superscriptaddress,longbibliography]{revtex4-1}%
\usepackage{graphicx}
\usepackage{float}
\usepackage{dcolumn}
\usepackage{bm,color}
\usepackage{amsmath,amssymb,dsfont,amstext,amsfonts}
\usepackage{extarrows}
\usepackage[colorlinks=true,linkcolor=blue,urlcolor=blue,citecolor=blue]{hyperref}
\usepackage{xcolor}
\usepackage{wasysym}
\usepackage{mathtools}
\usepackage{bbold}
\usepackage{verbatim}
\usepackage{tikz}
\usepackage[normalem]{ulem} 
\usepackage{color}

\def\mod{\:\mathrm{mod}\:}

\usepackage{pifont}

\makeatletter
\newcommand{\subalign}[1]{%
  \vcenter{%
    \Let@ \restore@math@cr \default@tag
    \baselineskip\fontdimen10 \scriptfont\tw@
    \advance\baselineskip\fontdimen12 \scriptfont\tw@
    \lineskip\thr@@\fontdimen8 \scriptfont\thr@@
    \lineskiplimit\lineskip
    \ialign{\hfil$\m@th\scriptstyle##$&$\m@th\scriptstyle{}##$\hfil\crcr
      #1\crcr
    }%
  }%
}
\makeatother

\usepackage{soul}

\begin{document}
\title{Sub-dimensional topologies, indicators and higher order phases}
\author{Gunnar F. Lange}
\thanks{Contributed equally}
\email{gfl25@cam.ac.uk}
\affiliation{TCM Group, Cavendish Laboratory, University of Cambridge, J.J.~Thomson Avenue, Cambridge CB3 0HE, United Kingdom}
\author{Adrien Bouhon}
\thanks{Contributed equally}
\email{adrien.bouhon@gmail.com}
\affiliation{Nordic Institute for Theoretical Physics (NORDITA), Stockholm, Sweden}
\affiliation{Department of Physics and Astronomy, Uppsala University, Box 516, SE-751 21 Uppsala, Sweden}
\author{Robert-Jan Slager}
\email{rjs269@cam.ac.uk}
\affiliation{TCM Group, Cavendish Laboratory, University of Cambridge, J.J.~Thomson Avenue, Cambridge CB3 0HE, United Kingdom}
\date{\today}
\begin{abstract}
The study of topological band structures have sparked prominent research interest the past decade, culminating in the recent formulation of rather prolific classification schemes that encapsulate a large fraction of phases and features. Within this context we recently reported on a class of unexplored topological structures that thrive on the concept of 
{\it sub-dimensional topology}. Although such phases have trivial indicators and band representations when evaluated over the complete Brillouin zone, they have stable or fragile topologies within sub-dimensional spaces, such as planes or lines. This perspective does not just refine classification pursuits, but can result in  observable features in the full dimensional sense. In three spatial dimensions (3D), for example, sub-dimensional topologies can be characterized by non-trivial planes, having general topological invariants, that are compensated by Weyl nodes away from these planes. As a result, such phases have 3D stable characteristics such as  Weyl nodes, Fermi arcs and edge states that can be systematically predicted by sub-dimensional analysis. Within this work we further elaborate on these concepts. We present refined representation counting schemes and address distinctive bulk-boundary effects, that include momentum depended  (higher order) edge states that have a signature dependence on the perpendicular momentum. As such, we hope that these insights might spur on new activities to further deepen the understanding of these unexplored phases.
\end{abstract}

\maketitle
\section{Introduction}
Topological effects in band structures has received significant attention in recent years~\cite{Rmp1,Rmp2}, leading to a myriad of topological phases and effects~\cite{Clas2,Clas1, codefects2, Nodal_chains, HolAlex_Bloch_Oscillations, ShiozakiSatoGomiK, Chenprb2012, discoverywti,regdef, kariyado2020selective,xiaoxing, UnifiedBBc, Wi2}. Of particular interest in this regard however is recent progress on mapping out a considerable fraction of topological materials. Namely, using constraints on band representations between high symmetry point in the Brillouin zone (BZ) that can reproduce the full K-theory in certain cases~\cite{Clas3}, elaborate schemes have emerged that upon comparing these condition in momentum space to real space provide for direct indicators of topological non-triviality ~\cite{Clas4, Clas5, mSI, mtqc,AdrienGunnarRobert2020}. This is usually phrased in terms of so-called symmetry indicators~\cite{Clas4} or elementary band representations~\cite{Clas5}. While the former is roughly obtained by considering the constraints in momentum space as a vector space, which delivers indicators upon dividing out Fourier transformed trivial atomic limit configurations, the rational behind considering elementary band representations (EBR) is that a split EBR must lead to a non-trivial behavior~\cite{Clas5}. That is, a topological configuration can by definition not be represented in terms of Wannier functions (of the localized kind) that also respect all symmetries.

This progress in itself has already sparked the discovery of new kinds of topologies. It was for example found that mismatches between stable symmetry indicators and split EBRs can be understood by the existence of fragile topological phases~\cite{Ft1}. Such fragile phases formally amount to a difference of trivial phases and can consequently be trivialized by the addition of {\it trivial} bands, rather than bands having opposite topological invariants \cite{bouhon2019wilson,Peri797,song2019fragile}. This concept of fragile topology in fact not only applies to symmetry indicated phases (i.e. those that can be deduced from the irreducible representation content at high symmetry points) but can be generalized by taking into account multi-gap conditions~\cite{bouhonGeometric2020}. Such phases can physically be understood as arising by braiding non-Abelian  frame charges \cite{Wu1273,BJY_nielsen, bouhon2019nonabelian}, leading to novel types of invariants and physical effects~\cite{Eulerdrive} . 

In recent work \cite{AdrienGunnarRobert2020}, we elucidated the idea of fragile topology in a magnetic context \cite{Axion1,Axion2,Axion3,Axion4,yang2021symmetryprotected, Axion5,Axion6,xiaoxing,mtqc} and outlined its connection to the concept of {\it sub-dimensional topology}. The essential idea articulates around the fact that while EBRs can be globally connected, thus appearing trivial in previous schemes, they can still be splittable on sub-dimenional spaces such as planes in 3D Brillouin zones. These sub-dimensional spaces can effectively be diagnosed in terms of symmetry indicators and EBR content. This is not merely an esoteric observation but results in real (in the 3D sense) consequences. The non-trivial plane might for example host stable invariants such as Chern numbers or spin Chern numbers that necessarily have to be compensated by Weyl nodes to lead to a globally connected EBR. As result, there are phases that have stable topological features (such as Chern planes and Weyl nodes) with distinct physical observables (such as edge states and Fermi arcs) that depend on this new concept, i.e the analysis of sub-dimensions that in turn reveal the necessary existence of nodes in the 3D band structure. While we presented an exhaustive table of all space groups in which the simplest form of this mechanism occurs~\cite{AdrienGunnarRobert2020}, it was conjectured that this new view can play a role in other settings as well.


In this work, we further develop the idea of fragile magnetic topology and sub-dimensional topology by investigating the bulk-corner and bulk-hinge correspondence, the twisted boundary conditions and the bulk-edge correspondence for these phases. This elucidates the connection between fragile magnetic topology, sub-dimensional topology and other well-studied phases such as higher-order topological insulators (HOTI). In particular, we explore the physical consequences of sub-dimensional topology by looking at the
hinge and edge state spectra of a sub-dimensional phase, and by examining its bulk-edge correspondence. Interestingly, we find that the sub-dimensional perspective can manifest itself by affecting the edge/hinge spectrum as function of the momentum along the hinge or direction perpendicular to the plane hosting the non-trivial sub-dimensional topology. We thus find as a main result that this refined perspective can results in specific consequences at the edge, momentum depended hinge spectra, thereby providing for new physical signatures.

The paper is organised as follows. In section \ref{sec:setup_section} we introduce the magnetic space-groups (MSG) and representation content which we will be considering. In section \ref{sec:corner_charges}, we present the spectra in various finite-dimensional geometries and comment on their connection to corner/edge charges and to HOTI. In section \ref{sec:RSI_TBC_main}, we further corroborate our findings by connecting them to the recently introduced idea of real-space invariants and twisted boundary conditions \cite{Twisted_BBC_theory, Peri797}. In section \ref{sec:Wilsonian_spectrum}, we finally connect our discussion to Wilson loops and a bulk-edge correspondence. We conclude in section \ref{sec:Conclusion}.

\section{Setup and topology for MSG75.5 and MSG77.18}\label{sec:setup_section}
\subsection{Setup and topology for MSG75.5}

\begin{figure}[ht!]
    \centering
    \includegraphics[width=\linewidth]{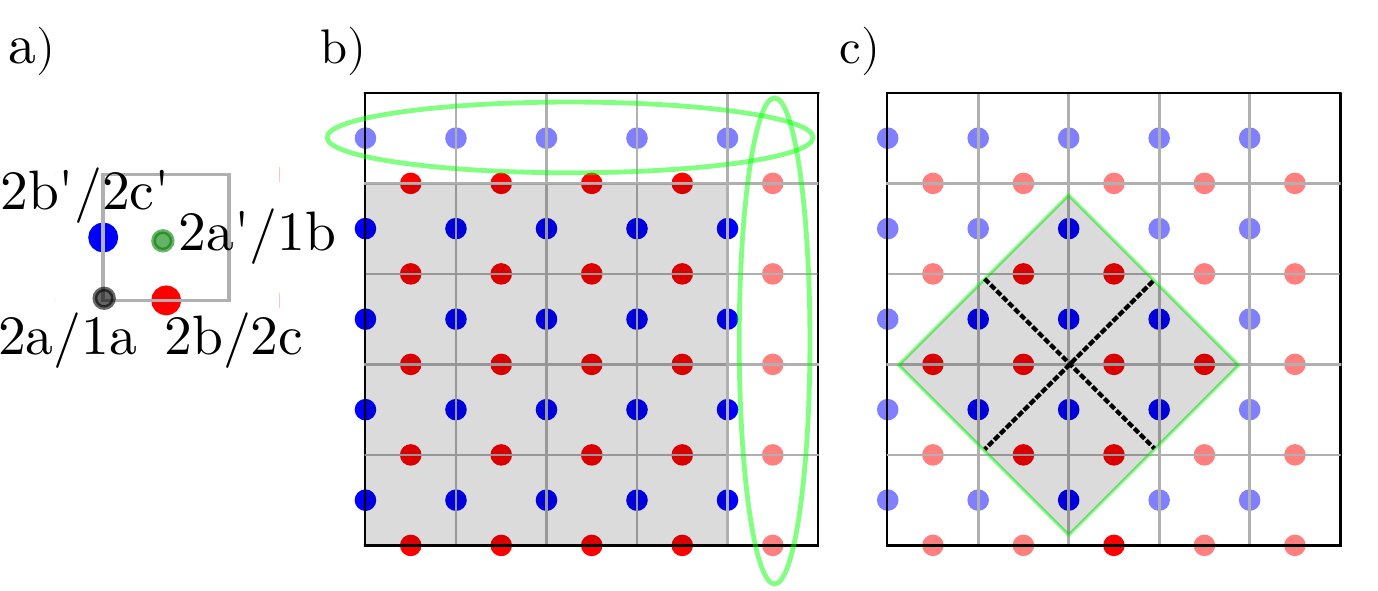}
    \caption{Illustration of the unit cell (panel a) and two possible lattice cuts that respect $C_4$ symmetry (panels b and c) for MSG75.5. a) The unit cell, with (m)WPs labelled, using both the convention of MSG75.5 ($2a, 2a', 2b, 2b'$) and of wallpaper group 10/p4 ($1a, 1b, 2c, 2c'$). We place one spin up and one spin down orbital at $2b$ and $2b'$ respectively. b) An integer number of units cells violates $C_4$ symmetry. This can be restored by considering a half-integer number of unit cells, which corresponds to ignoring the sites circled in green. c) To have $C_4$ symmetric sectors with no orbitals on the boundary, we can perform the cut shown in the green square, with the boundary between the sectors indicated.}
    \label{fig:75_5_setup}
\end{figure}

The magnetic space-group (MSG) 75.5 is generated from the tetragonal (non-magnetic) space group 75 (P4, generated by $C_4$ rotation) by including the antiunitary symmetry $(E|\tau)'$, with $E$ the identity, $\tau = \boldsymbol{a}_1/2+\boldsymbol{a}_2/2$, $(\cdot)'$ denoting time-reversal and $\boldsymbol{a}_i$ the primitive vectors of a (primitive) tetragonal Bravais lattice. Thus MSG75.5 is a Shubnikov type IV MSG which hosts anti-ferromagnetic ordering \cite{BradCrack}. We showed in \cite{AdrienGunnarRobert2020} that starting from magnetic Wyckoff position (mWP) $2b$, with sites $\boldsymbol{r}_A = \boldsymbol{a}_1/2$ and $\boldsymbol{r}_B = \boldsymbol{a}_2/2$, the real-space symmetries necessitate a minimum of four states to be present in the unit cell. Our choice of unit cell is shown in Fig.~\ref{fig:75_5_setup}a). We introduced a spinful model of this MSG with two sites per unit cell, each hosting two orbitals, in \cite{AdrienGunnarRobert2020} (summarized in Appendix \ref{ap:MSG75_5_model}). This model can be split into two disconnected two-band subspaces over the entire BZ whilst respecting all symmetries of MSG75.5, and thus realizes a split magnetic elementary band representation (MEBR) \cite{mtqc}. All symmetry indicators are trivial in our model, and therefore one of these subspaces necessarily realizes \textit{fragile} topology, i.e. it can be trivialized by coupling to trivial bands. The other two-band subspace realizes an atomically obstructed limit, where the electrons localize at a mWP distinct from the mWP of the atomic orbitals~\cite{Clas5}. We note, however, that both subspaces display non-trivial Wilson loop winding as discussed in section \ref{sec:Wilsonian_spectrum}. Thus whilst the obstructed atomic insulator label is useful pictorially, a more careful rigorous analysis based on the Wilson loop is needed in general, which we carry out in section \ref{sec:Wilsonian_spectrum}.  The split in our model can be written explicitly as:
\begin{equation*}
\mathrm{MEBR}^{2b}_{75.5} \rightarrow \underbrace{(\mathrm{MEBR}_{75.5}^{2b} \ominus \mathrm{MEBR}_{75.5}^{2a})}_{\mathrm{Lower\ subspace}}\oplus \underbrace{\mathrm{MEBR}_{75.5}^{2a}}_{\mathrm{Upper\ subspace}}.
\end{equation*}
Where $\ominus$ denotes formal subtraction of MEBRs. In terms of the spinful site-symmetry co-IRREPs (using notation from the \texttt{Bilbao Crystallographic Server} \cite{Bilbao2,Bilbao3,Bilbao}) this can be written as:
\begin{equation}
\label{eq:decomposition75.5}
    [\underbrace{(^1\overline{E}^2\overline{E})_{2b}\uparrow G \ominus (^1\overline{E}_1)_{2a} \uparrow G}_{\mathrm{Lower\ subspace}}] \oplus \underbrace{(^1\overline{E}_1)_{2a}\uparrow G}_{\mathrm{Upper\ subspace}}
\end{equation}
This decomposition can be determined directly from the momentum space IRREPs, using the formalism of (magnetic) topological quantum chemsistry \cite{Clas5,mtqc}. For MSG75.5, the unitary symmetries do not dictate any $k_z$ dependence. It is therefore legitimate to consider the planes containing the time-reversal invariant momentum points (TRIMPs) in the BZ separately, which results in an effective 2D model. We choose to focus on the plane $k_z = 0$. 

To explore the physical consequences of magnetic fragile topology in this system, we analyze its edge/corner spectrum and also consider its evolution under twisted boundary conditions (TBC). We therefore build a finite 2D lattice version of our model which respects $C_4$ symmetry. We consider two different ways of building this lattice, illustrated in Fig.~\ref{fig:75_5_setup} b)-c). As can be seen in Fig.~\ref{fig:75_5_setup}b), including an integer number of unit cells necessarily violates $C_4$ symmetry for a cut along the crystallographic axis. We therefore consider a half-integer number of unit cells in both direction for such a cut. We also consider how these spectra depend on the real-space termination, which leads us to also consider the cut in Fig.~\ref{fig:75_5_setup}c). This cut is also useful with regard to the TBC, which we discuss in section \ref{sec:RSI_TBC_main}, as there are no orbitals on the boundaries between regions related by $C_4$ symmetry. We refer to this as the diagonal cut.

Note that our finite model necessarily breaks the non-symmorphic antiunitary symmetry $(E|\tau)'$. This effectively reduces the symmetry to the spinful ferro/ferrimagnetic phase 75.1 (Shubnikov type I), which in the 2D case reduces to wallpaper group 10 (p4) with strong spin-orbit coupling (SOC) and without time-reversal symmetry (TRS). This group has been well-studied in e.g. \cite{CornerChargeSoC, Twisted_BBC_theory}. We reproduce some of their results to make the connection to the novel sub-dimensional topology in MSG77.18 more transparent.

Wallpaper group 10 (p4) has three (non-magnetic) WPs: $1a$ and $1b$ with point-group (PG) symmetry $C_4$ and $2c$ with PG symmetry $C_2$. This labeling is also shown in Fig.~\ref{fig:75_5_setup}a). This is similar to the model considered in \cite{Twisted_BBC_theory}, however, there the fragile phase originates from TRS. In our case, the fragile phase results from magnetic symmetries which manifest in the gluing of IRREPs in momentum space. In terms of the site-symmetry IRREPs of MSG75.1, our model realizes the decomposition in equation (\ref{eq:IRREP_decomposition_MSG_751}),
\begin{equation}\label{eq:IRREP_decomposition_MSG_751}
    \underbrace{[(^1\overline{E})_{2c} (^2\overline{E})_{2c}\ominus (^1\overline{E}_1)_{1a}(^2\overline{E}_1)_{1b}]}_{\mathrm{Lower \ subspace}}\oplus\underbrace{(^1\overline{E}_1)_{1a}(^2\overline{E}_1)_{1b}}_{\mathrm{Upper\ subspace}}.
\end{equation}

As suggested by Eq.~(\ref{eq:decomposition75.5}) and Eq.~(\ref{eq:IRREP_decomposition_MSG_751}), in the following we refer to the unoccupied band subspace as the \textit{obstructed phase} because it is formally compatible with an obstructed atomic limit in terms of its IRREPs content, and we call the occupied band subspace the \textit{fragile phase} because it can only be written as a subtraction of two atomic limits.

\subsection{Setup and topology for MSG77.18}
As there are no symmetry constraints in the $k_z$ direction for MSG75.5, we can freely consider topologies on the planes $k_z = 0$ and $k_z = \pi$ independently. This is not the case in MSG77.18, as was discussed in \cite{AdrienGunnarRobert2020}. This MSG is similar to MSG75.5, except that all $C_4$ rotations are replaced by screw rotations $C_{4_2} = (C_4|00\frac{1}{2})$, and the non-symmorphic time-reversal is replaced by $(E|\tau_d)'$ with $\tau_d = \boldsymbol{a}_1/2+\boldsymbol{a}_2/2+\boldsymbol{a}_3/2$. The non-symmorphic screw symmetry imposes connectivity constraints along the $k_z$ direction which prevent us from globally gapping the band structure in the 3D BZ. However, we can still gap the band structure on the planes $k_z = 0$ and $k_z = \pi$. We constructed a model in \cite{AdrienGunnarRobert2020}, summarized in appendix \ref{ap:MSG77_18_model}, which realizes a gapped band structures on these planes. Our model is based on mWP $2a$ in MSG77.18, with sites $\boldsymbol{r}_A  = \boldsymbol{a}_1/2+z\boldsymbol{a}_3$, $\boldsymbol{r}_B = \boldsymbol{a}_2/2+(z+1/2)\boldsymbol{a}_3$, where we specialize to $z = 0$. This model agrees with the model for MSG75.5 on the plane $k_z = 0$, and therefore everything we find for MSG75.5 holds on the plane $k_z = 0$ of MSG77.18 as well. 

To satisfy the symmetry constraints, we must necessarily have two nodal points along the $\overline{\Gamma\text{Z}}$ line and two other nodal points on the $\overline{\text{MA}}$ line, resulting in a semi-metallic phase as studied by e.g. \cite{Wi2,bouhon2017bulk,global_top_semi,BzduConversion,BbcWeyl,BJYNogaoseDSM,pseudographene,Zou2019,Wieder2020,PhysRevB.99.041301,PhysRevB.98.241103}. These Weyl nodes manifest through Fermi arcs in the surface spectrum, and as gapless states in the hinge spectrum as we show in Fig.~ \ref{fig:77_18_kz}. Furthermore, the pair of Weyl points on one vertical line are related by $C_2T$ symmetry and must thus have equal chirality, say positive chirality for the Weyl points on $\overline{\Gamma\text{Z}}$ \cite{AdrienGunnarRobert2020}. We can thus define a vertical cylinder surrounding the $k_z$-axis over which the Chern number must be $+2$. By the Nielsen-Ninomiya theorem this must in turn be compensated by the Weyl points appearing on the $\overline{\text{MA}}$ line contributing a Chern number of $-2$. We thus see that upon employing the sub-dimensional analysis we can {\it predicatively} enumerate in-plane non-triviality and, accordingly, necessarily present Weyl nodes that are the consequence of the in-plane topology \cite{AdrienGunnarRobert2020}. In the rest of this paper, we further explore the consequences of these topological structures in MSG75.5 and MSG77.18.

\section{Corner charges}\label{sec:corner_charges}
To elucidate the physical consequences of magnetic fragile phases in sub-dimensional topologies, we begin by considering the corner charges present in the system. Corner charges were studied in detail, for 2D non-magnetic systems, in \cite{CornerChargesFirst, CornerChargeSoC}. They can be related to electric multipole moments, as described in \cite{SOT2}. For an obstructed atomic or fragile insulator it can happen that charge neutrality and space-group symmetry are incompatible. Symmetry then necessitates an imbalance of ionic and electronic degrees of freedom, which gives rise to a filling anomaly $\chi$, defined as:
\begin{equation}
\chi = (\#\mathrm{Electronic\ sites}-\#\mathrm{Ionic\ sites}) \ \mathrm{mod}\ n
\end{equation}
Symmetry-allowed perturbations of the edges and corners can change this number in general, but it is always well-defined modulo some number $n$ related to the order of the symmetry. For the corner charges in MSG75.5 and MSG77.18, $n = 4$ due to the fourfold rotation symmetry.

If the edges are insulating, then the excess charge must localize at the corners of the system. If $\chi$ is incompatible with the order of the symmetry, then there will be fractional charges at the corners. This charge on the corner is only well-defined in the absence of an edge state, as edge states generically allow charge to flow away from the corner. Corner charges are therefore closely linked with higher-order topological insulators (HOTIs), as explored in \cite{Schindler2018,Codefects1,Wieder2020,PhysRevB.99.041301,CornerChargesFirst, Khalaf_sum_indicators, VanMiert2018,khalaf2019boundaryobstructed}. We find that this allows for a counting procedure to determine the excess charge in MSG75.5. We only consider half-filling, so that every ionic and electronic site contributes a single charge.

\subsection{Corner charges in MSG75.5}
\begin{figure}[ht!]
    \centering
    \includegraphics[width=\linewidth]{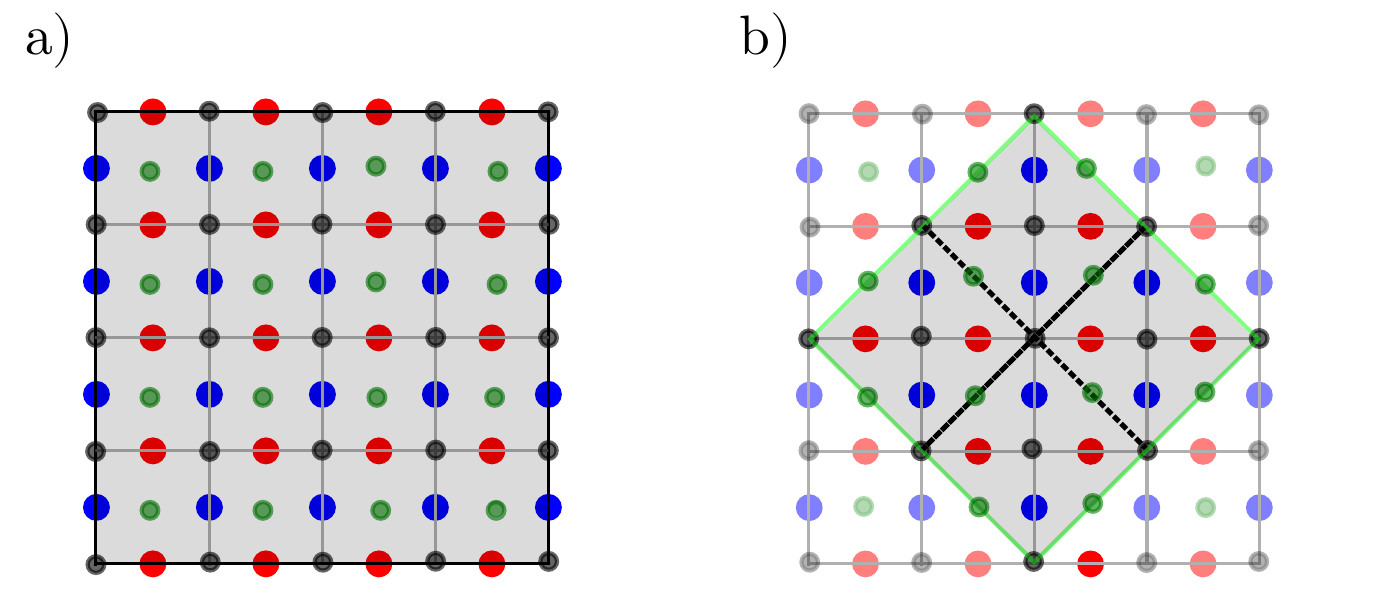}
    \caption{Same cuts for the $k_z = 0$ plane of MSG75.5/MSG77.18 as in Fig.~\ref{fig:75_5_setup}b) and c), but including the WPs where the electrons localize. The electrons localize on the smaller (black/green) sites, but the ions sit on the larger (red/blue) sites.}
    \label{fig:75_5_charge_counting}
\end{figure}

\begin{figure}[ht!]
    \centering
    \includegraphics[width=\linewidth]{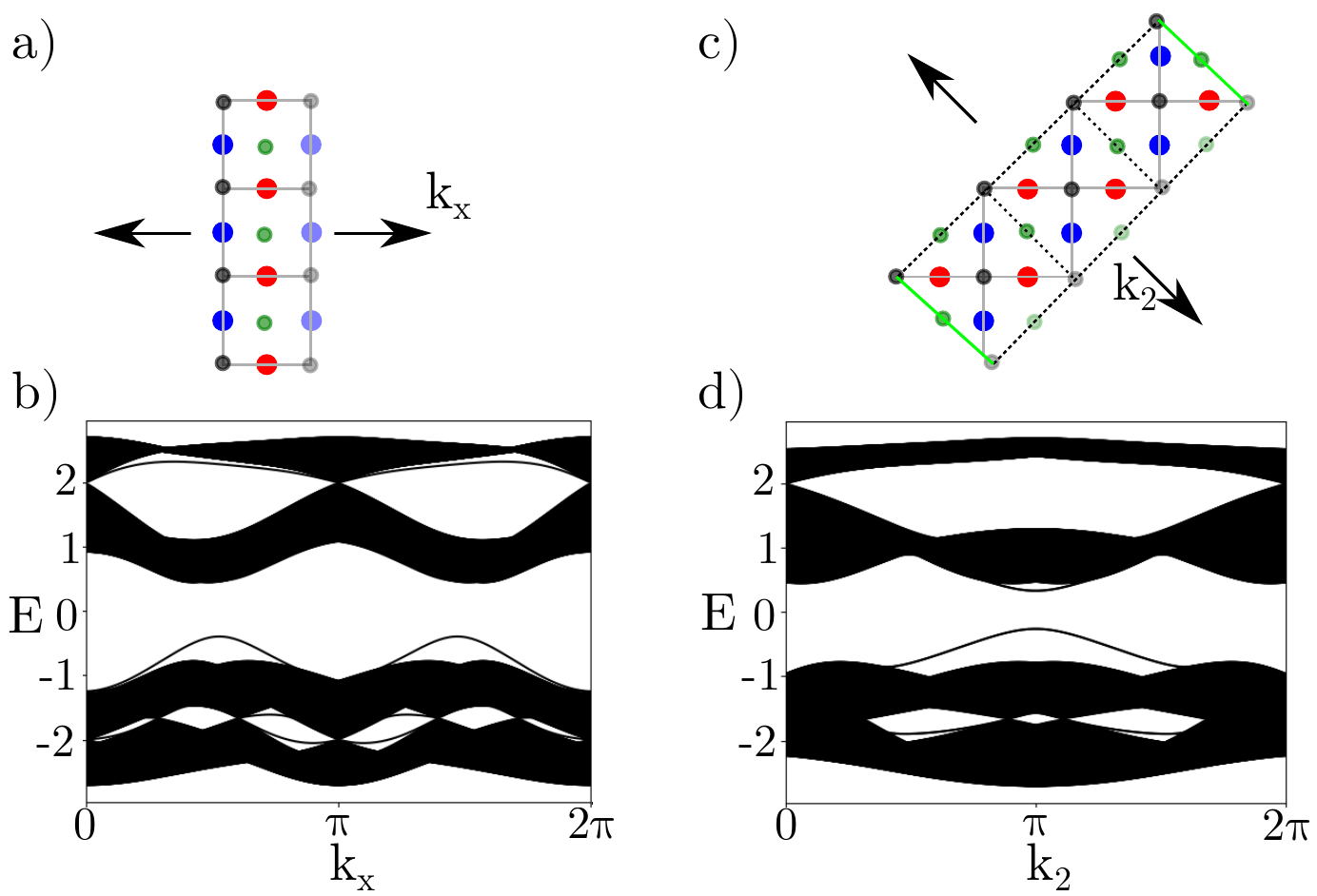}
    \caption{Counting of charges on the edge for MSG75.5. The arrows indicate which directions are considered to be periodic. The greyed out orbitals stem from adjacent unit cells and the black dotted lines indicate a possible choice of cell for the diagonal case. This is shown for the straight cut in panels a) and b) and the diagonal cut in panel c) and d), with associated edge band structures. The edge bands were calculated using the \texttt{PythTB} package \cite{PythTB2016}, with 150 unit cells perpendicular to the edge.}
    \label{fig:75_5_edge_counting}
\end{figure}

As $\mathrm{MEBR}_{75.5}^{2b}$ corresponds to a trivial insulator, with the electrons (center of band charges) localized at the ionic sites, its total filling anomaly must be zero. It follows that the filling anomalies of the occupied and unoccupied subspaces of the split EBR in MSG75.5 must sum to zero (modulo 4). We can therefore determine the filling anomaly of the fragile phase (i.e.~the occupied subspace) at half-filling by studying the atomic obstructed phase (i.e.~the unoccupied subspace) and counting the ionic and electronic sites in the system (this only works because of the vanishing bulk polarization, see \cite{CornerChargeSoC}). For this purpose, we redraw Fig.~\ref{fig:75_5_setup} with the sites of the magnetic WP$2a$ included. This corresponds to the non-magnetic WPs $1a$ and $1b$ and is shown in Fig.~\ref{fig:75_5_charge_counting}.
Counting the total number of electronic and ionic sites gives for the cut in Fig.~\ref{fig:75_5_charge_counting}a) a filling anomaly of $\chi = 41-40 = 1$. Note, however, that whether or not we include the electronic sites on the boundary is a matter of convention, as they only represent the localization centers of the electrons (they are not real sites in our model). We can therefore discount charges on the edge, but must do so in a $C_4$ symmetric fashion, as discussed above. We thus expect the obstructed phase to have an excess charge of $1$ mod $4$ electrons, and the fragile phase must then have a compensating excess charge of $3$ mod $4$ electrons. The same counting gives for the diagonal cut in \ref{fig:75_5_charge_counting}b) an anomaly of $\chi = 25-16 = 9$, which gives the same $\chi$ of $1$ mod $4$ in the obstructed phase.\\
To determine whether or not this filling anomaly gives rise to corner states, we must determine whether there is any excess charge localized on the edge of the system, which could result in a conducting edge. We show the counting of charges on the edge in Fig.~\ref{fig:75_5_edge_counting}, together with the edge spectra for the relevant cuts.
To count the orbitals, we must take care not to overcount sites which are periodic images of each other (indicated by greyed out orbitals). This gives a filling anomaly of $\chi = 7-7 = 0$ for Fig.~\ref{fig:75_5_edge_counting}a), and $14-12=2$ for Fig.~\ref{fig:75_5_edge_counting}b). Note, however, that we can remove charges on the edge, as long as we remove them symetrically from both edges, e.g. the edge charges are quantized mod 2. We therefore do not expect any fractional charges on the edge for either cut, and therefore expect quantized corner charges. We noted in Ref.~\cite{AdrienGunnarRobert2020}, that we have a quantized Berry phase of $\pi$ in the fragile subspace. This does not give rise to a topological edge state because the ionic sites are shifted from the origin by the same amount as the electronic sites, as discussed in section \ref{sec:Wilsonian_spectrum}. If this relation is violated, we expect an edge state to arise. We show the effect of removing various orbitals on the edge on the edge spectrum in Fig.~\ref{fig:ap_edge_termination_75_5} in Appendix \ref{ap:alternative_edge_terminations}. This confirms that we can get in-gap states by removing edge orbitals. When we remove pairs of orbitals at one edge, there are no topological in-gap edge states.

\begin{figure}[ht!]
    \centering
    \includegraphics[width=\linewidth]{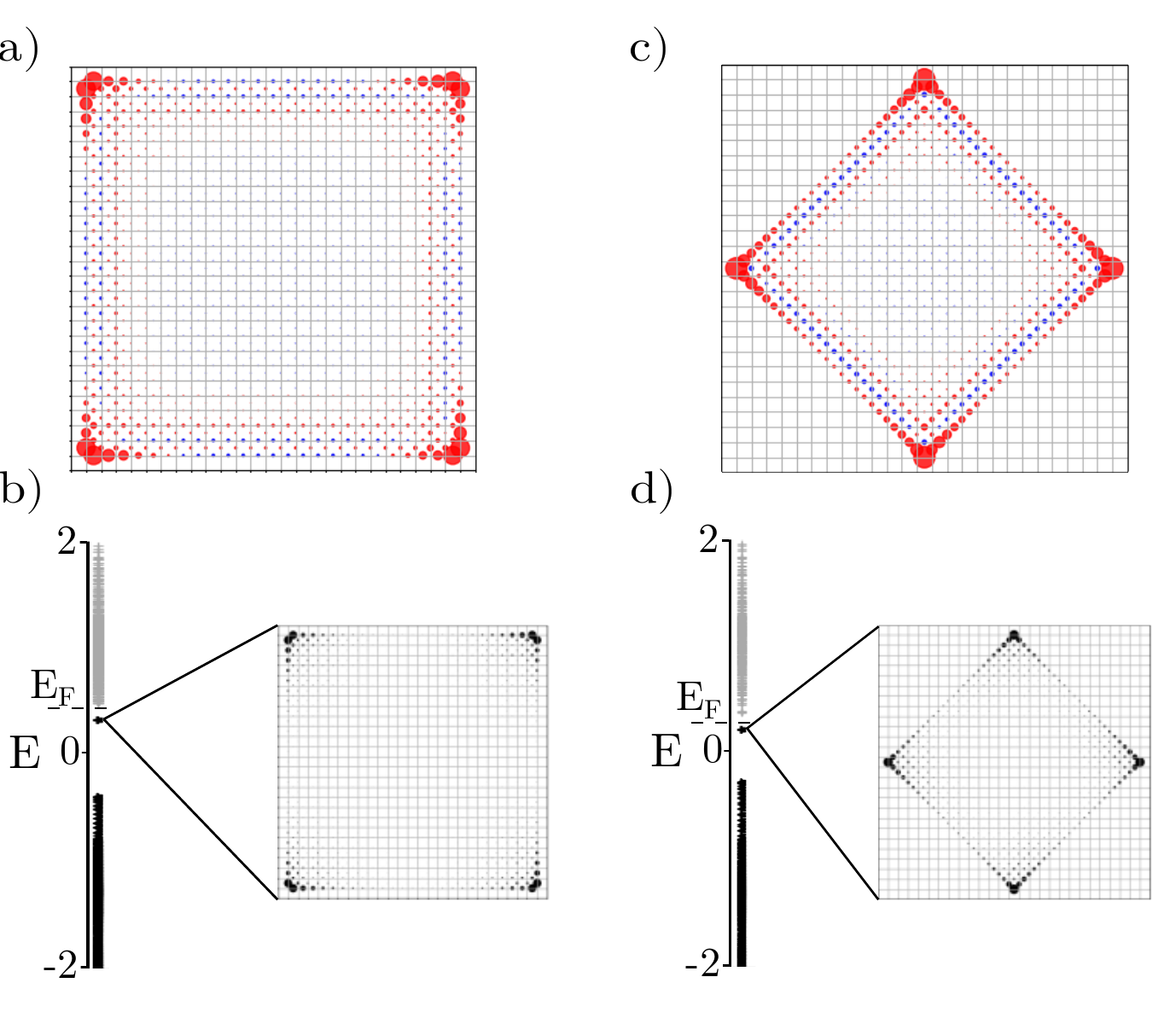}
    \caption{Corner charges and spectrum for the straight cut a) and b) and diagonal cut c) and d) respectively for MSG75.5. On the top, we show the charge distribution, with Fermi level set at the top of the gap, above the corner states. Red indicates excess charge (relative to the center), blue a deficit of charge where we sum over all occupied states. We confirm numerically that the excess electronic charge in a region away from the center is $3$ for both cuts. In the lower panels we show the spectrum, including the absolute value squared of a typical corner state. In both cuts, there are $N/2-1$ occupied states, $4$ corner states and $N/2-3$ unoccupied states, where $N$ is the total number of states. All calculations were done using the \texttt{PythTB} package \cite{PythTB2016}.}
    \label{fig:75_5_edge_charge}
\end{figure}

We note that this counting can also be used to predict the split of states into the occupied/unoccupied space in this case. If there are $N$ total states in the system, we expect $N/2-1$ states below the gap, $4$ corner states in the gap and $N/2-3$ states above the gap. Note, however, as shown in Fig.~\ref{fig:75_5_edge_counting}c) and d), that there are residual, model-specific, edge branches in the system which slightly extend into the gap but are not genuine topological in-gap states. These likely come from a nearby symmetry which our model only breaks weakly. We therefore expect some residual charge on the edges, but this charge is not topologically protected. To compute the corner charges, we therefore sum over the charge from the occupied bands contained in a region with finite thickness including the edge. We show the total charge for both cuts in Fig.~\ref{fig:75_5_edge_charge}, together with the associated spectra and typical corner states.
To confirm that these are indeed corner charges, we also plot the same system with a single spin removed on the boundary in Fig.~\ref{fig:ap_75_5_edge_charge_spin_removed} in Appendix \ref{ap:remove_single_spin}. This violates the integer quantization of charges (since we are removing one orbital while we remain at half-filling) and naturally leads to the appearance of an edge state with edge charges. In Fig.~\ref{fig:ap_75_5_edge_charge_spin_removed}, we clearly see that the edge states dominate the corner charges, illustrating that, in contrast, Fig.~\ref{fig:75_5_edge_charge} displays corner charges. We also confirm numerically that the excess charge in a region including the boundary is $3$ when fixing the Fermi energy above the corner states but below the upper (conduction) subspace. The sub-dimensional topology in MSG75.5 thus hosts corner charges. As the unitary symmetries do not dictate the $k_z$ dependence for MSG75.5, we expect that these $C_4$ symmetry protected corner charges are present for all values of $k_z$, leading to hinge states. 

\subsection{Corner charges in MSG77.18}
\begin{figure}[ht!]
    \centering
    \includegraphics[width=\linewidth]{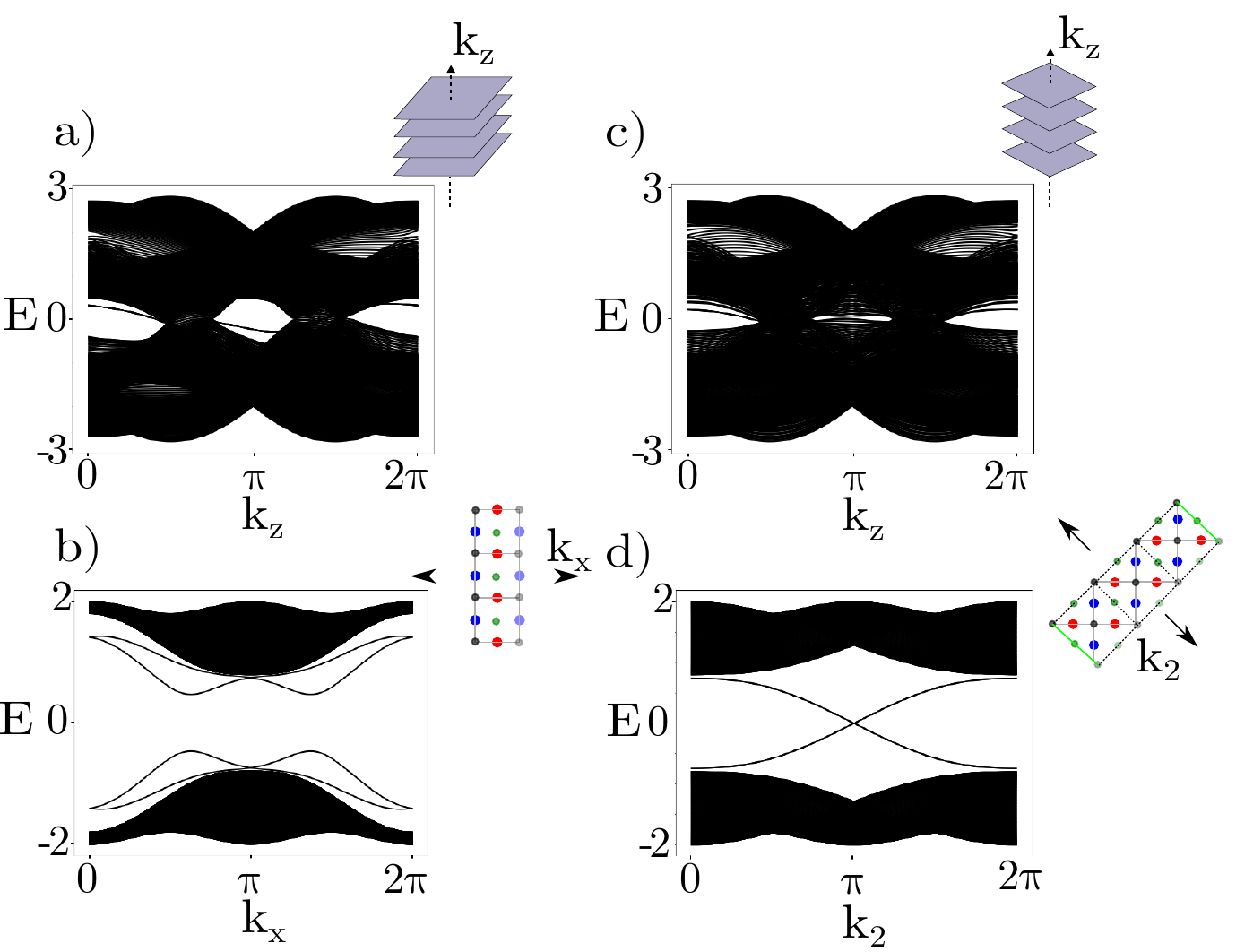}
    \caption{Hinge spectrum and edge spectrum in the $k_z = \pi$ plane for a slab calculation of MSG77.18. Panels a) and b) correspond to straight cuts. Panels c) and d) present data for the diagonal cut. All calculations were done using the \texttt{PythTB} package \cite{PythTB2016}}
    \label{fig:77_18_kz}
\end{figure}

\begin{figure}[ht!]
    \centering
    \includegraphics[width=\linewidth]{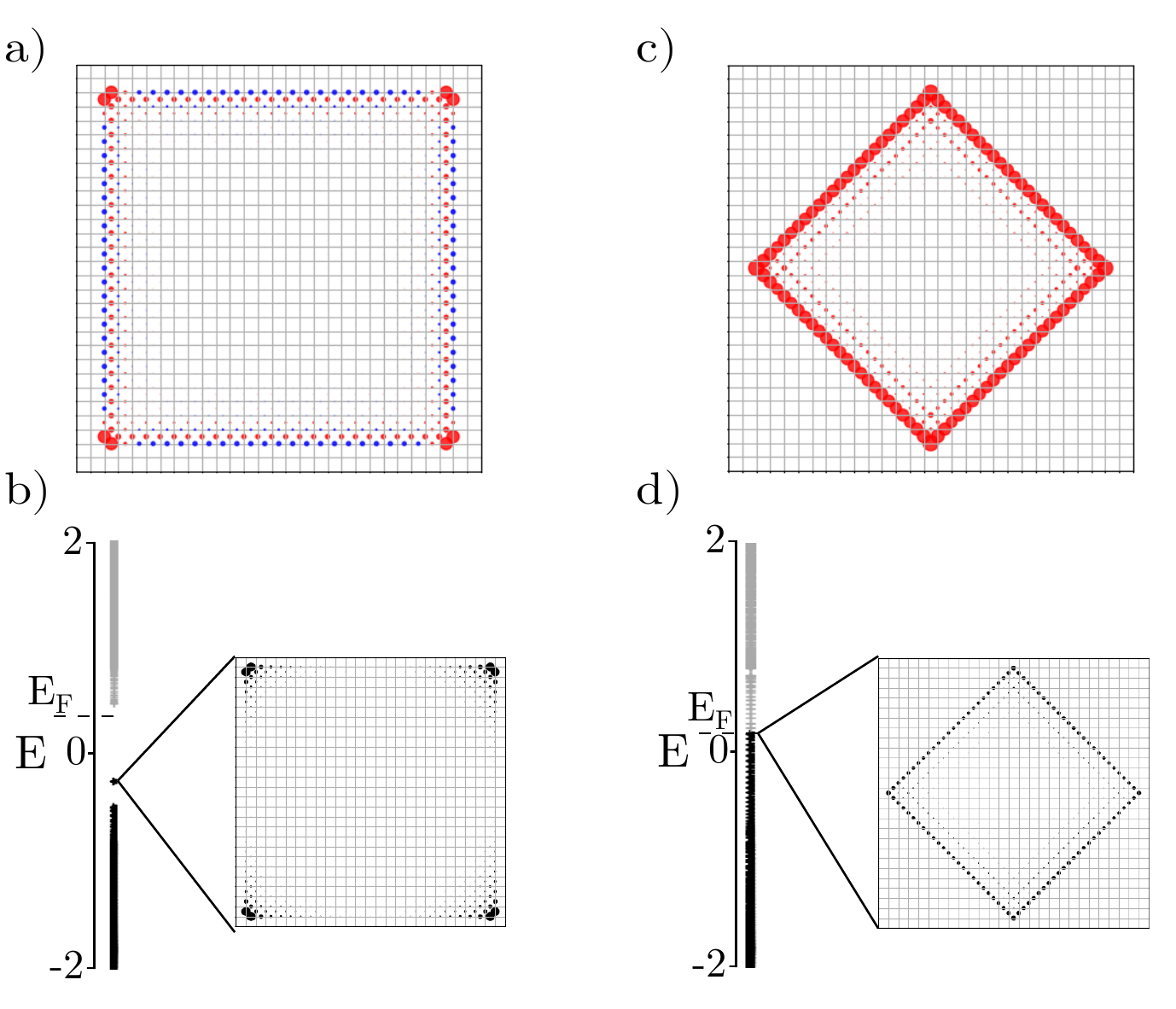}
    \caption{Corner charges for the straight cut a) and b) and diagonal cut c) and d) respectively for the $k_z = \pi$ plane of MSG77.18. On the top, we show the charge distribution, with Fermi level fixed to the same value as for MSG75.5, shown in Fig. \ref{fig:75_5_edge_charge}. Red indicates excess charge (relative to the center), blue a deficit of charge. We confirm numerically that the excess electronic charge in a region away from the center is $1$ for the straight cut. In the lower panels we show the spectrum, including the absolute value squared of a typical corner state. In the straight cut, there are $N/2-3$ occupied states, $4$ corner states, and $N/2-1$ empty states, with $N$ the total number of states.  All calculations were done using the \texttt{PythTB} package \cite{PythTB2016}.}
    \label{fig:77_18_edge_charge}
\end{figure}

The translational symmetry in the $z$-direction for MSG75.5 allows for the existence of hinge states that can be traced to the corner states of the fragile topology at $k_z = 0$ and $k_z = \pi$. The screw symmetry in MSG77.18 breaks this translational symmetry, and connects the planes at $k_z = 0$ and $k_z = \pi$. Our model for MSG77.18 was introduced in \cite{AdrienGunnarRobert2020}, and is discussed in Appendix \ref{ap:MSG77_18_model}. It agrees with the model for MSG75.5 on the plane $k_z = 0$, and we therefore expect corner charges on this plane. As we move along the $k_z$ direction, the screw symmetry necessitates the existence of Weyl nodes along $\overline{\Gamma \mathrm{Z}}$ and $\overline{\mathrm{MA}}$, which leads to Fermi arcs in the surface spectrum. We denote the $k_z>0$ coordinates of these Weyl nodes as $ k_{\overline{\Gamma \mathrm{Z}}}$ and $ k_{\overline{\mathrm{MA}}} $. In the hinge spectrum, we then expect a gap closing for all $k_z \in \big[\mathrm{min}\big( k_{\overline{\Gamma \mathrm{Z}}}, k_{\overline{\mathrm{MA}}}\big), \mathrm{max}\big( k_{\overline{\Gamma \mathrm{Z}}}, k_{\overline{\mathrm{MA}}}\big) \big]$. 
We plot the hinge spectrum for the straight and diagonal cut in Fig.~\ref{fig:77_18_kz}. 
We note that the symmetry $k_z \rightarrow -k_z$ is not maintained for edge/corner states, as the non-symmorphic $C_2T$ symmetry (relating $k_z$ to $-k_z$) is broken at the hinge. We also note that we get clear in-gap states for the hinge spectrum in the diagonal cut. We show the equivalent of Fig.~\ref{fig:75_5_edge_charge} for the $k_z = \pi$ plane of 77.18 in Fig.~\ref{fig:77_18_edge_charge}.
We note that the diagonal cut actually hosts edge modes on this plane, and that the corner charge differs from that on the $k_z = 0$ plane. This illustrates that the counting approach, which we adopted for MSG75.5 (or equivalently the $k_z = 0$ plane of MSG77.18), breaks down on the plane $k_z = \pi$. Thus care must be taken when extending counting schemes in 2D such as those discussed in \cite{CornerChargesFirst,CornerChargeSoC} to full 3D models. This difference between the planes can be understood by considering more carefully where the electrons localize on the plane $k_z = \pi$ of MSG77.18, as we describe in section \ref{sec:Wilsonian_spectrum}. The counting for the $k_z = 0$ plane of MSG77.18 only works because we have a direct map to an explicit 2D model, where the obstructed limits are known.

\section{Real space invariants and twisted boundary conditions}\label{sec:RSI_TBC_main}
Having explored the corner and edge spectrum, we will turn to relating the bulk and boundary features from a Wilson flow perspective in the next Section. However, before we move to this main topic, we shortly comment on the connection with the recently introduced concepts of twisted boundary conditions (TBC) \cite{Twisted_BBC_theory,Peri797} and real-space invariants. In particular, we note that on a single plane (e.g. $k_z = 0$) of MSG75.5, everything we have discussed is exclusively protected by $C_4$ symmetry. As a result, we can directly relate to the results as discussed in Ref. \cite{Twisted_BBC_theory}, providing an alternative perspective on fragile phases. This also serves to connect our work to other recent works, and can be relevant in an experimental setting \cite{Peri797}.
This formalism is most transparent when the phase being considered is gapless and when there are no orbitals on the boundary between regions related by $C_4$ symmetry. We therefore focus exclusively on the $k_z = 0$ plane of MSG75.5/MSG77.18, using the diagonal cut illustrated in Fig.~\ref{fig:75_5_setup}c), as MSG77.18 hosts a gapless phase on the $k_z = \pi$ plane for this cut.

\begin{figure}[ht!]
    \centering
    \includegraphics[width=\linewidth]{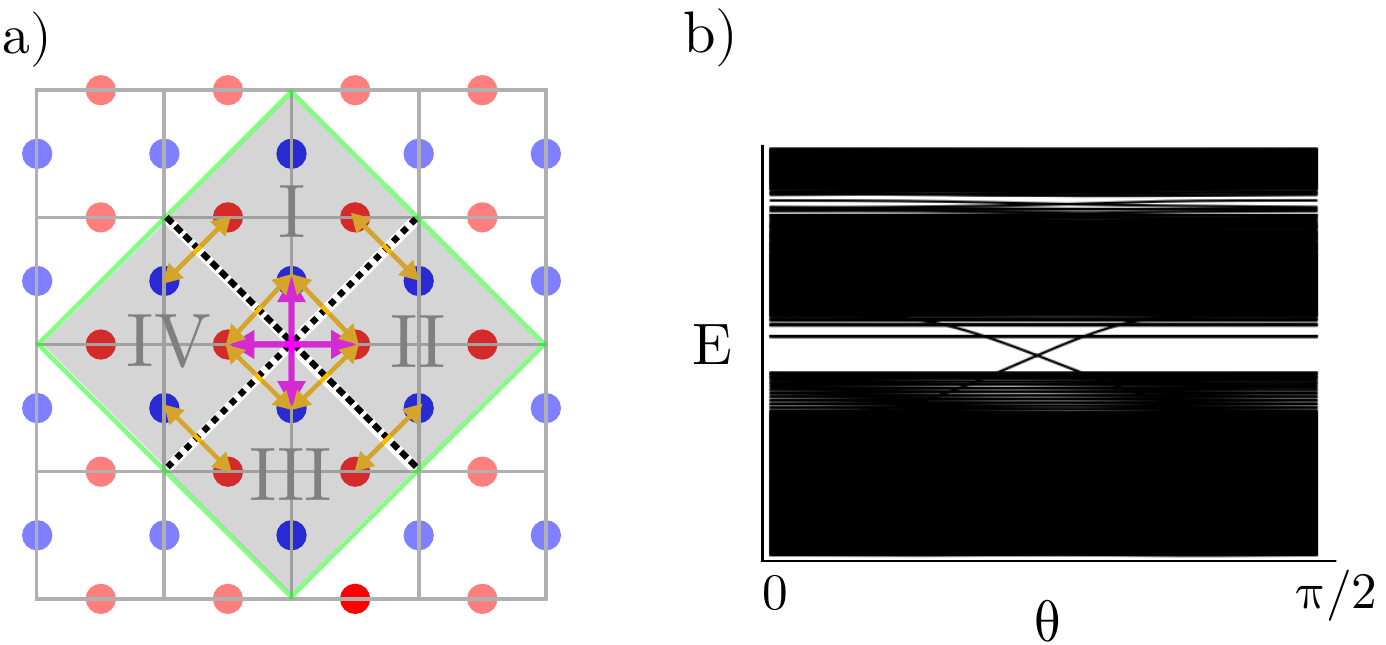}
    \caption{Twisted boundary conditions for a finite model of the $k_z = 0$ plane of MSG75.5. The hoppings between adjacent regions (orange arrows) in the clockwise/counterclockwise direction are multiplied by $\lambda = e^{\pm i\theta}$. The hoppings between diagonal regions (purple arrows) are multiplied by Re($\lambda^2$). }
    \label{fig:TBC_combined}
\end{figure}

This plane hosts wallpaper group p4 with SOC but without TRS. Using the expressions for the real-space invariants (RSIs) found in \cite{Twisted_BBC_theory}, we find that both the occupied and the unoccupied subspace have nonzero RSIs, the expressions of which can be found in appendix \ref{ap:RSI/TBC_updated}. This implies a non-trivial flow under twisted boundary conditions (TBC), where the coupling between the $C_4$ symmetric sectors is twisted by a factor $\lambda = e^{i\theta}$. The spectrum under this twisting is shown in Fig. \ref{fig:TBC_combined}, and we check that it agrees with the flow predicted from the RSIs. Note also the presence of the corner states which, as they are not cut by the TBCs, do not flow.
The RSIs can also be used to predict whether or not a set of bands are fragile. We find, as anticipated from the above discussion, that the lower band subspace is fragile, whereas the upper subspace is not. We refer for further details to Appendix \ref{ap:RSI/TBC_updated}.

\section{Bulk-edge correspondence}\label{sec:Wilsonian_spectrum}

The Wilson loop spectrum provides a powerful tool for the bulk-edge correspondence, which we here consider for the different edge terminations and at the different $k_z$-planes. We recall that upon integrating the Wilson loop along the momentum direction ($k_{\perp}$) perpendicular to the edge direction ($x_{\parallel}$), the Wilson phases give the $x_{\perp}$-component of the center of charge of the occupied band states, thus relating to a specific WP, which can then be used to predict an atomic obstruction (i.e.~the displacement of band charges to a WP distinct from the WP of the atomic orbitals) \cite{Zak2} and, subsequently, an edge-specific charge anomaly \cite{SSH}. Importantly, this argument is conclusive only when the Wilson loop spectrum is quantized as an effect of symmetries, since different WPs only then relate to topologically distinct phases (thus avoiding the adiabatic transfer of charges from one WP to an other). In the following we make use of the $\{0,\pi\}$ Berry phases (given as the sum of all Wilson loop phases) protected by $C_2T$ symmetry \cite{bouhon2019wilson} and the $\mathbb{Z}_2$ polarization protected by TRS and $C_2$ symmetry \cite{Ortix_Z2pol} associated with Kramers degeneracies of the Wilson loop spectrum. 

We start with a discussion of the effect of the symmetries of MSG75.5 and MSG77.18 on the Wilson loops in relation to (i) the sub-dimensional bulk topologies ($k_z=0,\pi$), and (ii) the two possible choices of geometries (straight versus diagonal, see Fig.~\ref{fig:Wilson_loop_77_18} a)). This allows us to determine which topological invariant ($\boldsymbol{Z}_2$ Berry phase and $\boldsymbol{Z}_2$ polarization) is associated with an edge geometry, as well as its actual value indicated by symmetry. We then motivate the bulk-edge correspondence for each case. 

\subsection{Topological insights from Wilson loop}\label{sec:Wilsonian_spectrum_insights}

We show in Fig.~\ref{fig:Wilson_loop_77_18} the Wilson loop spectrum for MSG77.18 (and MSG75.5, see below) corresponding to a straight geometry at the plane b) $k_z=0$ and d) $k_z=\pi$, and corresponding to the $45^{\circ}$-rotated (diagonal) geometry at the plane c) $k_z=0$ and e) $k_z=\pi$. Fig.~\ref{fig:Wilson_loop_77_18} a) shows the directions of integration in the momentum space with a full-line double arrow (blue) for the straight geometry and a dashed double arrow (orange) for the diagonal geometry. The Wilson loop spectrum for MSG75.5 is qualitatively the same as in Fig.~\ref{fig:Wilson_loop_77_18} b) for the straight geometry and Fig.~\ref{fig:Wilson_loop_77_18} c) for the diagonal geometry, both for the planes $k_z=0$ and $k_z=\pi$. Indeed, for MSG75.5 the sub-dimensional topologies at $k_z=0$ and $k_z=\pi$ are the same.  

\begin{figure}[t]
    \centering
    \includegraphics[width=\linewidth]{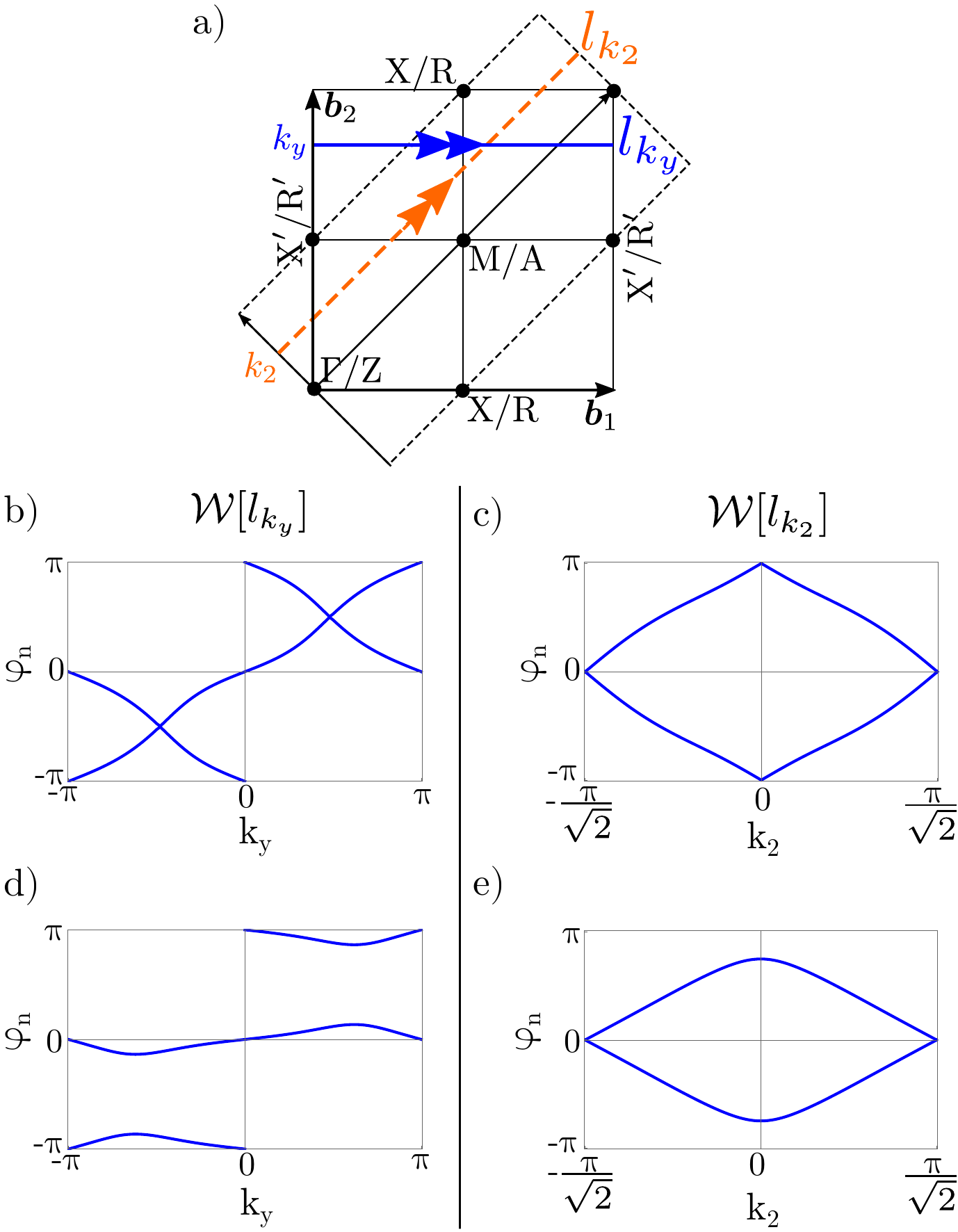}
    \caption{a) Planes of the Brillouin zone at $k_z=0$ ($\Gamma$, X, M, X') and $k_z=\pi$ (Z, R, A, R'), along straight directions $\{\boldsymbol{b}_1,\boldsymbol{b}_2\}$, and diagonal directions $\{\boldsymbol{b}_1+\boldsymbol{b}_2,-\boldsymbol{b}_1/2+\boldsymbol{b}_2/2\}$. b)-e) Wilson loop spectrum for MSG77.18 integrated (b,d) along the straight direction $\boldsymbol{b}_1$ (while varying $k_y$), and (c,e) along the diagonal direction $\boldsymbol{b}_1+\boldsymbol{b}_2$ (while varying $k_2$), for the plane (b,c) $k_z = 0$ and (d,e) $k_z=\pi$. The respective paths of Wilson loop integration for the straight geometry is $l_{k_y} = [(0,k_y)+\boldsymbol{b}_1\leftarrow (0,k_y)]$ (dashed orange), and for the diagonal geometry is $l_{k_2} = [(0,k_2)+\boldsymbol{b}_1+\boldsymbol{b}_2\leftarrow (0,k_2)]$ (full blue), shown as double arrows in a).}
      \label{fig:Wilson_loop_77_18}
\end{figure}

\subsubsection{Sub-dimensional topologies: $k_z=0$ vs $k_z=\pi$}

Let us now explain the differences between the Wilson loops in terms of the sub-dimensional topologies, i.e.~$k_z=0$ versus $k_z=\pi$, for MSG77.18. The sub-dimensional topology of the $k_z=0$ plane for MSG77.18 is identical to sub-dimensional topology of MSG75.5, with the high-symmetry points $\{\Gamma,\text{M}\}$ as TRIMPs (time reversal invariant momentum points) and with $[C_2T]^2=+1$ which protects the complete winding of two-band Wilson loop, thus indicating a nontrivial Euler class (fragile) topology \cite{bouhon2019wilson, AdrienGunnarRobert2020}. The winding of Wilson loop does not depend on the geometry (straight or diagonal) and we accordingly find complete Wilson loop windings in Fig.~\ref{fig:Wilson_loop_77_18} b) and c). At $k_z=\pi$, the sub-dimensional topology for MSG77.18 is characterized by the TRIMPs $\{\text{R},\text{R'}\}$ (which differ from the TRIMPs $\{\text{Z},\text{A}\}$ for MSG75.5) and with $[C_2T]^2=-1$ which implies the Kramers degeneracy of the energy bands over the whole momentum plane, while discarding Euler class topology \cite{AdrienGunnarRobert2020}. Accordingly, we find that there is no complete winding of the Wilson loops in Fig.~\ref{fig:Wilson_loop_77_18} d) and e).

\subsubsection{Symmetries and quantizations of Wilson loop: straight vs diagonal geometry}

We now turn to a detailed account of the quantizations and symmetries of Wilson loop spectra protected by symmetries for the two geometries, i.e.~straight versus diagonal. Our aim is to determine which topological invariant ($\boldsymbol{Z}_2$ Berry phase and $\boldsymbol{Z}_2$ polarization) can be associated with an edge geometry and when it is symmetry indicated (i.e.~with a definite value). These results combine the effects of $C_2$ symmetry, and the non-symmorphic anti-unitary symmetries TRS and $C_2T$. The derivation of the constraints on the Wilson loop due to the non-symmorphic anti-unitary symmetries (TRS and $C_2T$) is given in Appendix \ref{WL_sym}. Below we write $\mathcal{W}[l_{k}]$ the Wilson loop with the spectrum (phases) $\{\varphi_1(k),\varphi_2(k)\}$ evaluated over the non-contractible base path $l_k$ that is parametrized by $k=k_y$ in the straight geometry, and by $k=k_2$ in the diagonal geometry [Fig.~\ref{fig:Wilson_loop_77_18} a)].



In the straight geometry, the Wilson loop phases at $k_y=0$ and $k_y=\pi$ are quantized to $(\varphi_1,\varphi_2)=(0,\pi)$ by the $C_2$ symmetry, i.e. it follows from the $C_2$ eigenvalues of the band eigenstates at the high-symmetry points $\Gamma$ and M \cite{Wi1,Bouhon_global,bouhon2019wilson}, i.e.~the Berry phase must be $\gamma_B[l_{k_y=0,\pi}] = \varphi_1+\varphi_2 \mod 2\pi  =\pi \mod 2\pi$. This is true on both planes $k_z=0,\pi$. Then on the $k_z=0$ plane, by the $\boldsymbol{Z}_2$ quantization of the Berry phase protected by $C_2T$ ($[C_2T]^2=+1$), we conclude that the Berry phase must be $\pi$ for all $k_y$. On the $k_z=\pi$ plane, we have $[C_2T]^2=-1$ and there is the question of the quantization of the Berry phase. We have shown in Appendix \ref{WL_sym} that the non-symmorphic $C_2T$ imposes $\varphi_2(k_y) = -\varphi_1(k_y) + \pi\mod\ 2\pi$ on the spectrum of the Wilson loop. We thus get $\gamma_B(k_y) = \varphi_1(k_y)+\varphi_2(k_y) = \pi\mod 2\pi$ for all $k_y$. We note that the non-symmorphic TRS furthermore requires $\varphi_2(-k_y) = \varphi_1(k_y) + \pi\mod\ 2\pi$ (see Appendix \ref{WL_sym}) which explains the global structure of the Wilson loops in \ref{fig:Wilson_loop_77_18} b) and d). Also, since there is no path $l_{k_y}$ for which TRS squares to $-1$ for all $k_x$, no $\boldsymbol{Z}_2$ polarization can be defined in the straight geometry. 


In the diagonal geometry, the Wilson loop is quantized by $C_2$ at $k_2=\pm\pi/\sqrt{2}$ to $(\varphi_1,\varphi_2)=(0,0)$, for $k_z=0$ and $k_z=\pi$ (this follows from the fact that the $C_2$ eigenvalues of the occupied-band eigenstates at X and at R are all equal to $\mathrm{i}$ or $-\mathrm{i}$), such that the Berry phase is zero. (At $k_2=0$, and $k_z=0$ or $k_z=\pi$, on the contrary, there is no quantization of the Wilson loop from $C_2$.) At $k_z=0$, the $\boldsymbol{Z}_2$ quantization of the Berry phase (protected by $C_2T$ with $[C_2T]^2=-1$) implies it must be zero for all $k_2$, in agreement with Fig.~\ref{fig:Wilson_loop_77_18} c). At $k_z=\pi$, the non-symmorphic $C_2T$ symmetry requires $\varphi_2(k_2)=-\varphi_1(k_2)$ (see Appendix \ref{WL_sym}), hence again the Berry phase must be zero for all $k_2$, in agreement with Fig.~\ref{fig:Wilson_loop_77_18} e). We now address the effect of non-symmorphic TRS which requires $\varphi_2(k_2)=\varphi_1(-k_2)$ for $k_z=0$ and $k_z=\pi$ (see Appendix \ref{WL_sym}), which explains the global structure of the Wilson loops in Fig.~\ref{fig:Wilson_loop_77_18} c) and e). Furthermore, from the square of TRS, $T^2 = -\mathrm{e}^{-\mathrm{i} (k_x+k_y+k_z)}$, we predict one diagonal per $k_z$-plane along which $T^2=-1$. We find $T^2=-1$ for $k_x=k_y\mod2\pi$ at $k_z=0$, i.e.~on the diagonal $\overline{\Gamma \text{M}}$ (connecting the TRIMPs $\Gamma$ and M), and $T^2=-1$ for $k_x=k_y+\pi \mod 2\pi$ at $k_z=\pi$, i.e.~on the diagonal $\overline{\text{RR'}}$ (connecting the TRIMPs R and R'). This results in the presence of Kramers degeneracies of the Wilson loop spectrum at $k_2=0$ for $k_z=0$, namely $(\varphi_1,\varphi_2)=(\pi,\pi)$, and at $k_2=\pm \pi/\sqrt{2}$ for $k_z=\pi$, namely $(\varphi_1,\varphi_2)=(0,0)$. Moreover, the combination of $C_2$ and TRS leads to the definition of a $\nu[l]\in\boldsymbol{Z}_2$ polarization on these diagonals which is directly indicated by the phases of the Kramers degeneracies, i.e.
\begin{equation}
    \begin{aligned}
        \nu^{(k_z=0)}[l_{k_2=0}] &= 1,\\
        \mathrm{and}~\nu^{(k_z=\pi)}[l_{k_2=\pi/\sqrt{2}}] &= 0 .
    \end{aligned}
\end{equation} 
We show below that these have important consequences for the bulk-edge correspondence.

\subsection{Sub-dimensional charge anomalies and edge states}

Let us first summarize the findings of the previous section. The Berry phase is $\pi$ in the straight geometry on both planes $k_z=0$ and $k_z=\pi$. The Berry phase is $0$ in the diagonal geometry on both planes $k_z=0$ and $k_z=\pi$. There is no $\boldsymbol{Z}_2$ polarization in the straight geometry. In the diagonal geometry, there a nontrivial $\boldsymbol{Z}_2$ polarization on $l^{(k_z=0)}_{k_2=0}=\overline{\Gamma\text{M}}$, and a trivial one on $l^{(k_z=\pi)}_{k_2=0}=\overline{\text{RR'}}$.

\subsubsection{Straight geometry}

We consider here the bulk-edge correspondence at $k_z=0$ for which the phases for MSG75.5 and MSG77.18 are the same. We use MSG75.5 to give a detailed account of the bulk-edge correspondence in that momentum plane. 

Let us start with a discussion of the straight edge geometry. Let us argue explicitly for MSG75.5 that the $\pi$-Berry phase does not indicate a charge anomaly. Indeed, the Wyckoff position $2b$ of the atomic orbitals and the Wyckoff positions $2a$ of the obstructed band charges both have, component-wise, one site centered in the unit cell and one site shifted to the unit cell boundary. I.e.~defining the (unordered) sets of component-wise positions relative to the unit cell center 
\begin{equation}
\begin{aligned}
\{\boldsymbol{r}^{(2b)}_{A,i},\boldsymbol{r}^{(2b)}_{B,i}\}  &= \{0,a/2\}\mod a,\\
\mathrm{and}~\{\boldsymbol{r}^{(2a)}_{C,i},\boldsymbol{r}^{(2a)}_{D,i}\}  &= \{0,a/2\}\mod a , 
\end{aligned}
\end{equation}
for $i=x,y$, there is no difference. Therefore, even in the case of an obstruction (from WP $2b$ to WP $2a$) there is no charge anomaly, i.e.~no mismatch between the number of charges localized at the distinct WPs, in a ribbon system with straight edge cuts and assuming that a slide of the ribbon contains an integer number of unit cells (see Appendix \ref{ap:alternative_edge_terminations} with the results for a fractional number of unit cells). This can also be readily checked through direct counting. We thus conclude that there is no topological edge branch along the straight edges. We can readily transpose this conclusion is directly for MSG77.18 at $k_z=0$, as well as at $k_z=\pi$ for MSG75.5. 

Since the Berry phase is also $\pi$ at $k_z=\pi$ for MSG77.18, we find similarly to $k_z=0$ that there is no edge state in the straight geometry (assuming an integer number of unit cells per slice of the ribbon). We conclude that in the straight geometry there no effect of the sub-dimensional topology (i.e.~comparing $k_0=\pi$ and $k_z=\pi$) on the bulk-edge correspondence for MSG77.18. 




\subsubsection{Diagonal geometry} 

To fix ideas, we assume that $x_1=x_{\perp}$ is the direction perpendicular to the edge, and $x_2=x_{\parallel}$ the direction parallel to it. Then, we define the diagonal unit cell through $\boldsymbol{a}_{\perp} = \boldsymbol{a}_1/2+\boldsymbol{a}_2/2 = a/\sqrt{2} \hat{e}_{1} $ and $\boldsymbol{a}_{\parallel}  = -\boldsymbol{a}_1+\boldsymbol{a}_2 = \sqrt{2}a \hat{e}_{2}$, such that invariance under a translation by $\boldsymbol{a}_{\parallel}$ is satisfied. Note that this unit cell is different from the edge unit cell in figure \ref{fig:77_18_kz}. Writing the atomic positions in the diagonal axes, i.e. $\boldsymbol{r} = (x_1,x_2)$, the perpendicular component ($x_1$) of the Wyckoff positions in the diagonal geometry are
\begin{equation}
\begin{aligned}
(\boldsymbol{r}^{(2b)}_{A,1},\boldsymbol{r}^{(2b)}_{B,1})  &= \dfrac{a}{2\sqrt{2}}(1,1)\mod \dfrac{a}{\sqrt{2}},\\
\mathrm{and}~(\boldsymbol{r}^{(2a)}_{C,1},\boldsymbol{r}^{(2a)}_{D,1})  &= (0,0)\mod \dfrac{a}{\sqrt{2}} .
\end{aligned}
\end{equation}
The zero Berry phase obtained for the diagonal geometry thus indicates that there must be an even number of charges at WP $2b$ and at WP $2a$. Assuming an obstruction of the charges from WP $2b$ to WP $2a$, the diagonal edge cut leads to a total charge anomaly of $\pm 2e$ for the ribbon if we assume that a slide of the ribbon contains an integer number of diagonal unit cells (see Appendix \ref{ap:alternative_edge_terminations} with the results for a fractional number of unit cells). This means that we have a charge anomaly of $\pm e$ per diagonal edge. 

Let us now make use of the $\nu\in\boldsymbol{Z}_2$ polarization, which we have seen is well defined in the diagonal geometry. At $k_z=0$, we have found $\nu_{k_2=0} = 1$ which corresponds to the Kramers degenerate Wilson loop phases $(\varphi_1,\varphi_2)=(\pi,\pi)$. This implies that the band charges are not obstructed, i.e.~they can be located at WP $2b$. There is thus no charge anomaly and following there is no edge state. This conclusion holds for MSG75.5 at $k_z=0$ and $k_z=\pi$, and for MSG77.18 at $k_z=0$ only. 

Considering now MSG77.18 at $k_z=\pi$, we have found $\nu_{k_2=\pi/\sqrt{2}} = 0$ which corresponds to the Kramers degenerate Wilson loop phases $(\varphi_1,\varphi_2)=(0,0)$. This implies an obstruction of the band charges, i.e.~shifted from the atomic mWP $2a$ of MSG77.18. We have argued above that with an obstruction there is a charge anomaly of $\pm e$ per diagonal edge (assuming an integer number of diagonal unit cells in one slice of the ribbon). We also know that the $\boldsymbol{Z}_2$ polarization predicts the presence of an odd number integer-valued electronic charge at a single edge \cite{Ortix_Z2pol}, which by virtue of the Kramers degeneracy means a \textit{half-integer-valued charge per spin}. The topological (helical) edge states take the form of an odd number of edge Kramers pairs per edge. This is fully consistent with the numerical results shown in Fig.~\ref{fig:77_18_kz} where we find one Kramers pair of edge states per edge. 
 
We finally conclude that there is a non-trivial effect of the sub-dimensional topology on the bulk-edge correspondence that is observable in the diagonal geometry.  

\section{Conclusion}\label{sec:Conclusion}
In this work we have further explored the concept of sub-dimensional topology. While these topologies have a connected EBRs and thus appear trivial, they feature split EBRs on sub-dimensional spaces, such as planes, in the Brillouin zone. These in-plane topologies are subsequently compensated by Weyl nodes and have full dimensional non-trivial features. We in particular find that these concepts can be related to more refined counting and symmetry indicator arguments. Most notably, however, is the connection of these insights to consequences on the edge. We find that the sub-dimensional topology results in distinctive bulk-boundary signatures. These include hinge spectra and edge states that have a distinct dependence on the perpendicular momentum, underpinning the physical significance of this new topological concept. We therefore hope that our results will result in the further exploration of these features.

\begin{acknowledgments}
 R.-J.~S. acknowledges funding from the Marie Sk{\l}odowska-Curie programme under EC Grant No. 842901 and the Winton programme as well as Trinity College at the University of Cambridge. G.F.L acknowledges funding from the Aker Scholarship.
\end{acknowledgments}

\bibliography{references}
\clearpage
\newpage
\appendix

\section{Summary of the models}
Here we reproduce the models used for MSG75.5 and MSG77.18. These models were originally introduced in \cite{AdrienGunnarRobert2020}, and are written in the Bloch basis:
\begin{equation}
    \vert \varphi_{\alpha,\sigma} , \boldsymbol{k}\rangle = \sum_{\boldsymbol{R}\in \boldsymbol{T}} \mathrm{e}^{\mathrm{i} \boldsymbol{k} \cdot (\boldsymbol{R}+\boldsymbol{r}_{\alpha})} \vert w_{\alpha,\sigma} , \boldsymbol{R}+\boldsymbol{r}_{\alpha} \rangle,
\end{equation}
Where $\alpha \in \{A,B\}$ labels the sites in the unit cell, and $\sigma \in \{\uparrow, \downarrow\}$ labels spin components. We choose to order our basis as $\boldsymbol{\varphi} = (\varphi_{A,\uparrow},\varphi_{A,\downarrow},\varphi_{B,\uparrow},\varphi_{B,\downarrow})$. A more thorough analysis of these models in momentum space, including band structures, can be found in \cite{AdrienGunnarRobert2020}.
\subsection{The model for MSG75.5}\label{ap:MSG75_5_model}
Our model for MSG75.5 is defined as:
\begin{equation}
\label{model_75_5}
\begin{aligned}
    &H(\boldsymbol{k}) = t_1 f_1(\boldsymbol{k}) \sigma_z\otimes\sigma_z \\
    &+ t_2 f_2(\boldsymbol{k}) \sigma_y\otimes\mathbb{1}+ t_3 f_3(\boldsymbol{k})   \sigma_x\otimes\mathbb{1} \\
    &+ \lambda_1 g_1(\boldsymbol{k}) \mathbb{1}\otimes\sigma_+ + \lambda_1^*   g_1^*(\boldsymbol{k}) \mathbb{1}\otimes\sigma_- \\
    &+
    \lambda_2 g_2(\boldsymbol{k}) \sigma_x\otimes\sigma_+ +
    \lambda_2^*  g_2^*(\boldsymbol{k}) \sigma_x\otimes\sigma_- ,
\end{aligned}
\end{equation}
with $\sigma_{\pm} = (\sigma_x \pm \mathrm{i} \sigma_y)/2$ and lattice form factors
\begin{equation}
    \begin{array}{ll}
        f_1  = \cos \boldsymbol{a}_1\boldsymbol{k} -\cos \boldsymbol{a}_2\boldsymbol{k} ,&
        g_1  = \sin \boldsymbol{a}_1 \boldsymbol{k} - \mathrm{i}\sin \boldsymbol{a}_2 \boldsymbol{k},\\
        f_2  = \cos \boldsymbol{\delta}_1 \boldsymbol{k}-
        \cos \boldsymbol{\delta}_2 \boldsymbol{k} ,
        & g_2  = \sin \boldsymbol{\delta}_1 \boldsymbol{k} - \mathrm{i}\sin \boldsymbol{\delta}_2 \boldsymbol{k} ,\\
         f_3  = \cos \boldsymbol{\delta}_1 \boldsymbol{k} + \cos \boldsymbol{\delta}_2 \boldsymbol{k}, &
    \end{array}
\end{equation}
These are defined in terms of the bond vectors $\boldsymbol{\delta}_{\left(\substack{1\\2}\right)} = (\boldsymbol{a}_1 \left(\substack{-\\+}\right) \boldsymbol{a}_2)/2$.  We have assumed that $\{t_1,t_2,t_3\}$ are real, while $\{\lambda_1,\lambda_2\}$ can be complex. We fix $t_1, t_2, t_3 = 1$, and $\lambda_1,\lambda_2 = (1/2) \mathrm{e}^{\mathrm{i} \pi/5} $. 

\subsection{The model for MSG77.18}\label{ap:MSG77_18_model}
Our model for MSG77.18 is defined by adding an extra term $H'$ to the above model for MSG75.5:
\begin{multline}
\label{eq:model_77}
    H'(\boldsymbol{k}) = H[f_1,f_2',f_3',g_1,g_2'](\boldsymbol{k}) + \\
    \rho_1 h_1(\boldsymbol{k}) \sigma_x \otimes \sigma_z + \rho_2 h_2(\boldsymbol{k}) \sigma_y\otimes \sigma_z ,
\end{multline}
Where $H(\boldsymbol{k})$ is given in Equation \ref{model_75_5}, and the lattice form factors have been extended to 3D momentum space,
\begin{equation}
    \begin{aligned}
        f'_2(\boldsymbol{k}) &= \left(\cos \boldsymbol{\delta}'_1\boldsymbol{k} - 
        \cos \boldsymbol{\delta}'_2\boldsymbol{k} +
        \cos \boldsymbol{\delta}'_3\boldsymbol{k} -
        \cos \boldsymbol{\delta}'_4\boldsymbol{k}
        \right)/2,\\
        f'_3(\boldsymbol{k}) &= \left(\cos \boldsymbol{\delta}'_1\boldsymbol{k} + 
        \cos \boldsymbol{\delta}'_2\boldsymbol{k} +
        \cos \boldsymbol{\delta}'_3\boldsymbol{k} +
        \cos \boldsymbol{\delta}'_4\boldsymbol{k}
        \right)/2,\\
        g'_2(\boldsymbol{k}) &= \left(\sin \boldsymbol{\delta}'_1\boldsymbol{k} - \mathrm{i} 
        \sin \boldsymbol{\delta}'_2\boldsymbol{k} -
        \sin \boldsymbol{\delta}'_3\boldsymbol{k} + \mathrm{i}
        \sin \boldsymbol{\delta}'_4\boldsymbol{k}
        \right)/2,\\
        h_1(\boldsymbol{k}) &= \left(\sin \boldsymbol{\delta}'_1\boldsymbol{k} + 
        \sin \boldsymbol{\delta}'_2\boldsymbol{k} +
        \sin \boldsymbol{\delta}'_3\boldsymbol{k} +
        \sin \boldsymbol{\delta}'_4\boldsymbol{k}
        \right)/2,\\
        h_2(\boldsymbol{k}) &= \left(\sin \boldsymbol{\delta}'_1\boldsymbol{k} - 
        \sin \boldsymbol{\delta}'_2\boldsymbol{k} +
        \sin \boldsymbol{\delta}'_3\boldsymbol{k} -
        \sin \boldsymbol{\delta}'_4\boldsymbol{k}
        \right)/2,
    \end{aligned}
\end{equation}
with $\boldsymbol{\delta}'_{1,2} =   \boldsymbol{\delta}_{1,2}+\boldsymbol{a}_3/2$, and $\boldsymbol{\delta}'_{3,4} = - \boldsymbol{\delta}_{1,2}+\boldsymbol{a}_3/2$, and with new real parameters $\rho_1,\rho_2 \in \mathbb{R}$. We fix $\rho_1=-1$ and $\rho_2=-2/5$.

\section{Symmetries of the Wilson loop due to non-symmorphic TRS and $C_2T$ symmetry}\label{WL_sym}

We here derive the effect of non-symmorphic TRS and $C_2T$ symmetry on the Wilson loop for the different geometries, i.e.~straight versus diagonal shown in Fig.~\ref{fig:Wilson_loop_77_18} a). We do it explicitly for MSG77.18 but the final results apply to MSG75.5 as well. Let us write $\mathcal{W}[l_{\boldsymbol{k}_0}] = \langle \boldsymbol{u} , \boldsymbol{k}_0+\boldsymbol{K} \vert \prod\limits_{\boldsymbol{k}}^{\boldsymbol{k}_0+\boldsymbol{K}\leftarrow\boldsymbol{k}_0} \mathcal{P}_{\boldsymbol{k}} \vert \boldsymbol{u} , \boldsymbol{k}_0 \rangle $ the Wilson loop over the occupied Bloch eigenvectors $\{\vert u_n , \boldsymbol{k}_0 \rangle\}_{n=1,2}$ integrated over the base loop $l_{\boldsymbol{k}_0} = [\boldsymbol{k}_0+\boldsymbol{K}\leftarrow\boldsymbol{k}_0]$ that crosses the Brillouin zone with $\boldsymbol{K}$ defined as the smallest reciprocal lattice vector in that direction. We define the anti-unitary representation of the non-symmorphic TRS for MSG77.18, $(E\vert \tau_d)' $ where $\tau_d = (\boldsymbol{a}_1+\boldsymbol{a}_2+\boldsymbol{a}_3)/2$, in the basis of the occupied Bloch eigenstates through $ \mathcal{T}(\boldsymbol{k}) = \langle \boldsymbol{\psi},-\boldsymbol{k} \vert ^{(E\vert \tau_d)'} \vert \boldsymbol{\psi},\boldsymbol{k} \rangle = \mathrm{e}^{\mathrm{i} \boldsymbol{k}\cdot \tau_d} \langle \boldsymbol{u},-\boldsymbol{k}\vert (\sigma_x\otimes -\mathrm{i}\sigma_y) \mathcal{K} \vert \boldsymbol{u},\boldsymbol{k}\rangle = \mathrm{e}^{\mathrm{i} \boldsymbol{k}\cdot \tau_d} \langle \boldsymbol{u},-\boldsymbol{k}\vert (\sigma_x\otimes -\mathrm{i}\sigma_y)  \vert \boldsymbol{u}^*,\boldsymbol{k}\rangle \mathcal{K} = \mathcal{U}\mathcal{K} $ where $\mathcal{K}$ is the complex conjugation and $\mathcal{U}$ is unitary \cite{AdrienGunnarRobert2020}. It is then convenient to write the unitary representation with the phase factor of the non-symmorphicity removed, i.e.~$U(\boldsymbol{k}) = \mathrm{e}^{-\mathrm{i} \boldsymbol{k}\cdot \tau_d} \mathcal{U}(\boldsymbol{k})$ (this will correspond below to taking the ``periodic gauge'' \cite{hourglass,Bouhon_global,bouhon2017bulk} in the Wilson loop over a non-contractible path of the torus Brillouin zone). Similarly to the symmorphic case \cite{Wi1}, the constraint imposed by $(E\vert\tau_d)'$ on the Wilson loop is found to be $ U^*(\boldsymbol{k}_0+\boldsymbol{K})^{-1} \mathcal{W}^{T}[l_{-\boldsymbol{k}_0-\boldsymbol{K}}] U^*(\boldsymbol{k}_0) =  \mathcal{W}[l_{\boldsymbol{k}_0}] $ with $l_{-\boldsymbol{k}_0-\boldsymbol{K}} = [-\boldsymbol{k}_0\leftarrow -\boldsymbol{k}_0-\boldsymbol{K}]$, which we rewrite as 
\begin{equation}
    \mathrm{e}^{-\mathrm{i} \boldsymbol{K}\cdot \tau_d} \mathcal{U}^*(\boldsymbol{k}_0+\boldsymbol{K})^{-1} \mathcal{W}^{T}[l_{-\boldsymbol{k}_0-\boldsymbol{K}}] \mathcal{U}^{*}(\boldsymbol{k}_0) = \mathcal{W}[l_{\boldsymbol{k}_0}] .
\end{equation}
We thus conclude that in the straight geometry, i.e.~taking $\boldsymbol{k}_0=(0,k_y)$ and $\boldsymbol{K} = \boldsymbol{b}_1$, the Wilson loop spectrum satisfies the following symmetry
\begin{equation}
       \left\{ \varphi_n[l_{(0,-k_y)}] \right\}_{n=1,2} = \left\{ \varphi_n[l_{(0,k_y)}]+\pi\mod 2\pi \right\}_{n=1,2} ,
\end{equation}
while in the diagonal geometry, i.e.~taking $\boldsymbol{k}_0=(0,k_2)$ and $\boldsymbol{K} = \boldsymbol{b}_1+\boldsymbol{b}_2$, the symmetry reads
\begin{equation}
       \left\{ \varphi_n[l_{(0,-k_2)}] \right\}_{n=1,2} = \left\{ \varphi_n[l_{(0,k_2)}] \right\}_{n=1,2} .
\end{equation}

We now consider the effect of the non-symmorphic $C_2T$ symmetry of MSG77.18, $(C_{2}\vert \tau_d)'$, on the Wilson loop spectrum. Its anti-unitary representation in the occupied Bloch eigenstates reads $\mathcal{A}(\boldsymbol{k}) =  \langle \boldsymbol{\psi},-C_2\boldsymbol{k} \vert ^{(C_{2}\vert \tau_d)'} \vert \boldsymbol{\psi},\boldsymbol{k} \rangle = \mathrm{e}^{\mathrm{i} C_2\boldsymbol{k}\cdot \tau_d} \langle \boldsymbol{u},-C_2\boldsymbol{k}\vert (\sigma_x\otimes \mathrm{i}\sigma_x)  \vert \boldsymbol{u}^*,\boldsymbol{k}\rangle \mathcal{K} = \mathcal{R} \mathcal{K}$ where $\mathcal{R}$ is unitary. Defining $R(\boldsymbol{k}) = \mathrm{e}^{-\mathrm{i} C_2\boldsymbol{k}\cdot \tau_d} \mathcal{R}(\boldsymbol{k})$, we then find, similarly to the case of TRS above, $ R^*(\boldsymbol{k}_0+\boldsymbol{K})^{-1} \mathcal{W}^*[l_{\boldsymbol{k}_0}] R^*(\boldsymbol{k}_0) =  \mathcal{W}[l_{\boldsymbol{k}_0}]$, which we rewrite as 
\begin{equation}
    \mathrm{e}^{-\mathrm{i} C_2\boldsymbol{K}\cdot \tau_d} \mathcal{R}^*(\boldsymbol{k}_0+\boldsymbol{K})^{-1} \mathcal{W}^*[l_{\boldsymbol{k}_0}] \mathcal{R}^{*}(\boldsymbol{k}_0) = \mathcal{W}[l_{\boldsymbol{k}_0}] .
\end{equation}
We therefore conclude that in the straight geometry ($\boldsymbol{k}_0=(0,k_y)$, $\boldsymbol{K} = \boldsymbol{b}_1$) the Wilson loop spectrum satisfies the following symmetry
\begin{equation}
       \left\{ \varphi_n[l_{(0,k_y)}] \right\}_{n=1,2} = \left\{ -\varphi_n[l_{(0,k_y)}]+\pi\mod 2\pi \right\}_{n=1,2} ,
\end{equation}
while in the diagonal geometry ($\boldsymbol{k}_0=(0,k_2)$, $\boldsymbol{K} = \boldsymbol{b}_1+\boldsymbol{b}_2$) the symmetry reads
\begin{equation}
       \left\{ \varphi_n[l_{(0,k_2)}] \right\}_{n=1,2} = \left\{ -\varphi_n[l_{(0,k_2)}] \right\}_{n=1,2} .
\end{equation}

We emphasize that these Wilson loop symmetries hold similarly at $k_z=0$ and at $k_z=\pi$ since $\boldsymbol{K}$ has no $k_z$ component. This is also the reason why these results also apply to MSG75.5. 

We conclude by noting that these results are fully consistent with the computed Wilson loops in Fig.~\ref{fig:Wilson_loop_77_18}. 

\section{Additional figures}
In this Appendix we present additional figures to further detail the findings presented in the main text. In particular, we present the edge spectrum which results from removing various orbitals on the edges in appendix \ref{ap:alternative_edge_terminations}, and we present the charge distribution when removing a single spin on the entire edge in appendix \ref{ap:remove_single_spin}.

\subsection{Alternative edge termination}\label{ap:alternative_edge_terminations}
\begin{figure}[ht!]
    \centering
    \includegraphics[width=\linewidth]{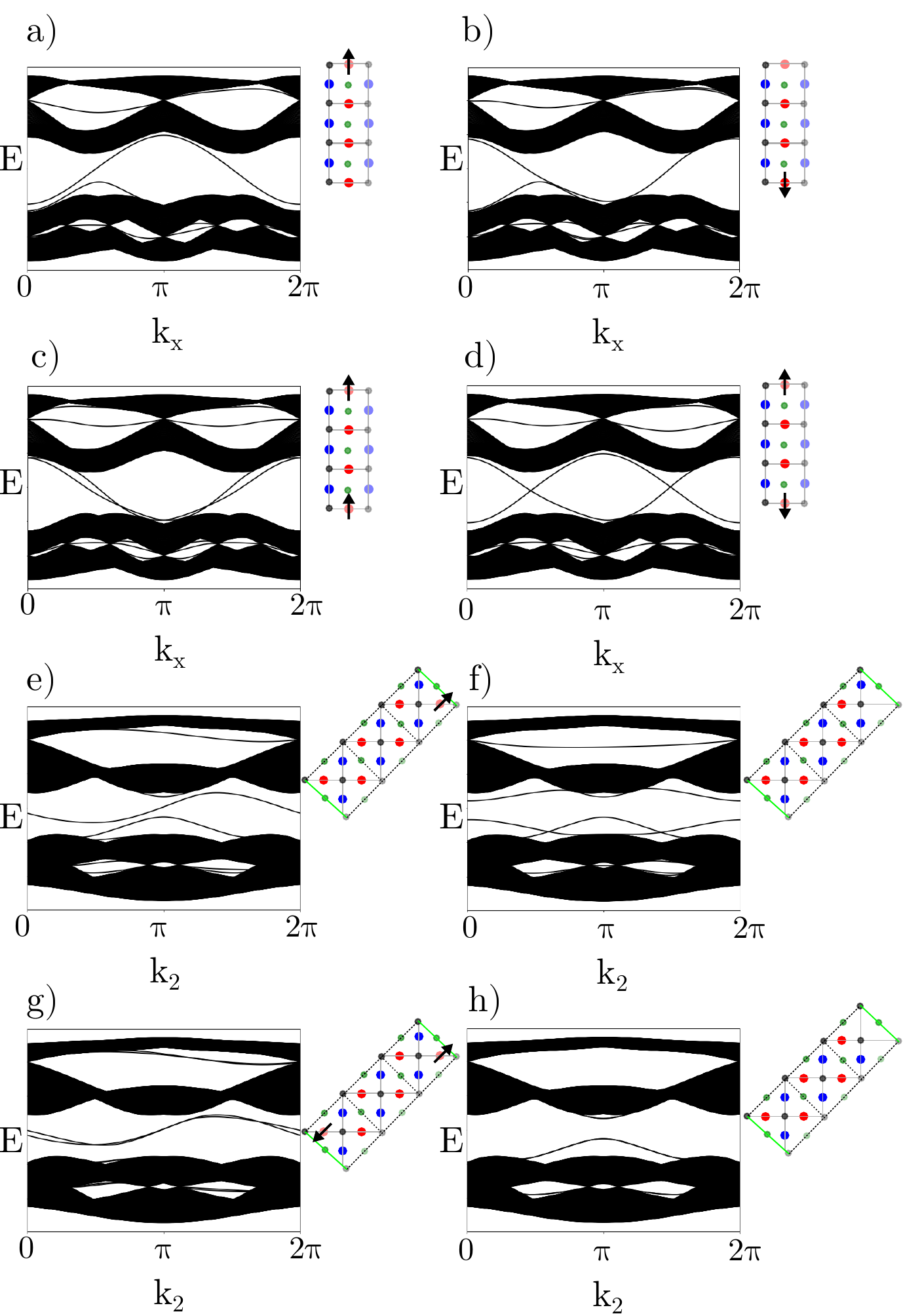}
    \caption{Edge spectra for a selection of removed edge orbitals in the straight cut (panel a - panel d) and diagonal cut (panel e - panel h) for MSG75.5/the $k_z = 0$ plane of MSG77.18. A full circle indicates the presence of both spin components, a single arrow indicates that the other spin has been removed and the lack of a site indicates that both spin components on this site have been removed. a) Removing a down spin from the top edge. b) Removing an up spin from the bottom edge. c) Removing a down spin from both edges. d) Removing a down spin from the top edge and an up spin from the bottom edge. e) Removing a down spin from the A site on the top edge. f) Removing both spin components from the A site on the top edge. g) Removing a down spin from the A site on the top edge and an up spin from the A site at the bottom edge. h) Removing all spins from the A and B site at the top edge. All in-gap states are singly degenerate. The states extending slightly into the gap in h) are doubly degenerate.}
    \label{fig:ap_edge_termination_75_5}
\end{figure}

\begin{figure}[ht!]
    \centering
    \includegraphics[width=\linewidth]{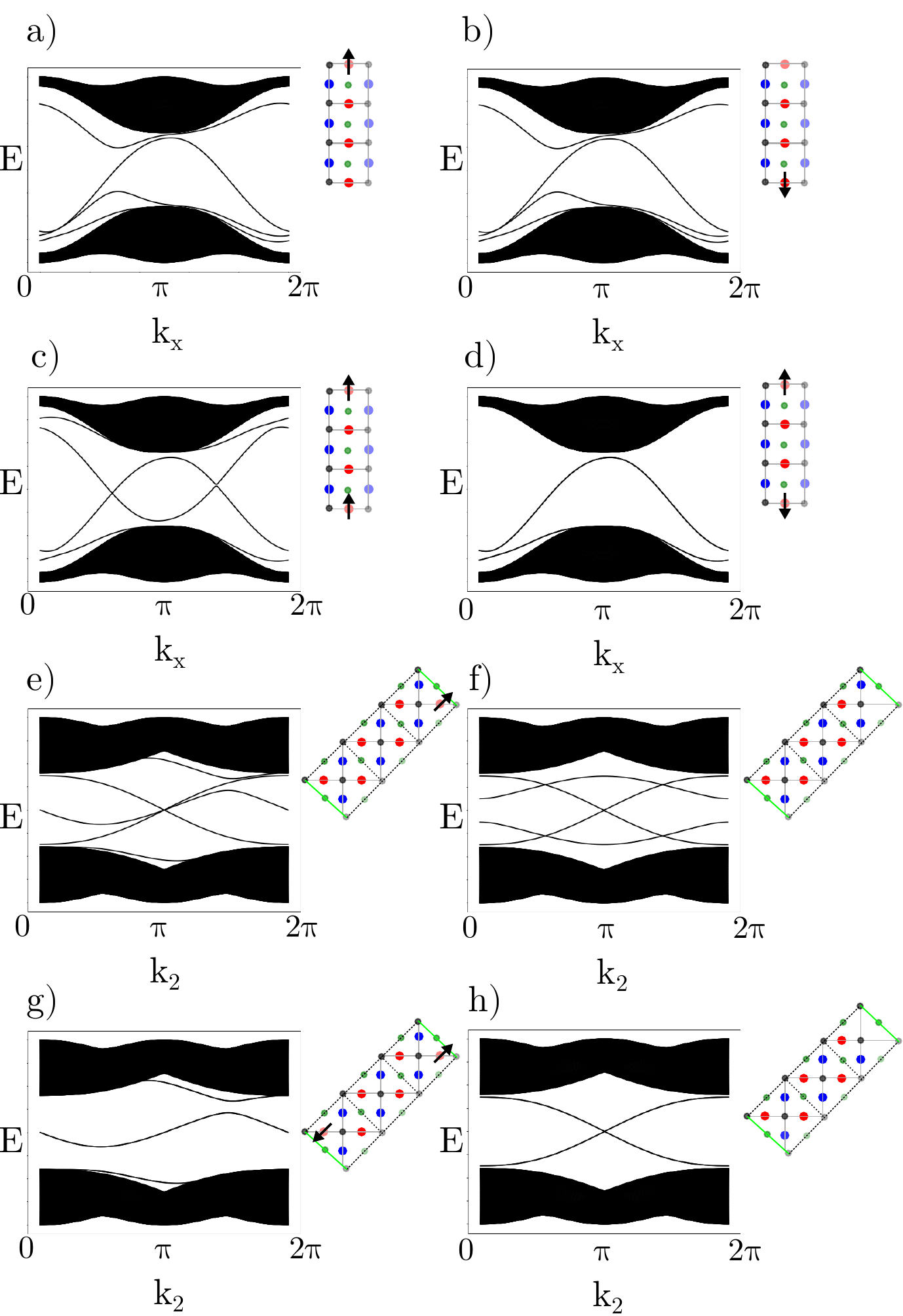}
    \caption{Edge spectra for a selection of removed edge orbitals in the straight cut (panel a - panel d) and diagonal cut (panel e - panel h) for the $k_z = \pi$ plane of MSG77.18. A full circle indicates the presence of both spin components, a single arrow indicates that the other spin has been removed and the lack of a site indicates that both spin components on this site have been removed.  a) Removing a down spin from the top edge. b) Removing an up spin from the bottom edge. c) Removing a down spin from both edges. d) Removing a down spin from the top edge and an up spin from the bottom edge. e) Removing a down spin from the A site on the top edge. f) Removing both spin components from the A site on the top edge. g) Removing a down spin from the A site on the top edge and an up spin from the A site at the bottom edge. h) Removing all spins from the A and B site at the top edge. The in-gap bands in d), g) and h) are doubly degenerate.}
    \label{fig:ap_edge_termination_77_18}
\end{figure}

We noted in Section \ref{sec:corner_charges} of the main text that removing a single orbital at the edge of the system can induce edge states. In Fig.~\ref{fig:ap_edge_termination_75_5} we show a selection of ways this can be done for MSG75.5 (equivalently the $k_z = 0$ plane of MSG77.18). In Fig.~\ref{fig:ap_edge_termination_77_18}, we show the corresponding plot for the $k_z = \pi$ plane of MSG77.18.


\subsection{Removing a single spin}\label{ap:remove_single_spin}
\begin{figure}[ht!]
    \centering
    \includegraphics[width=\linewidth]{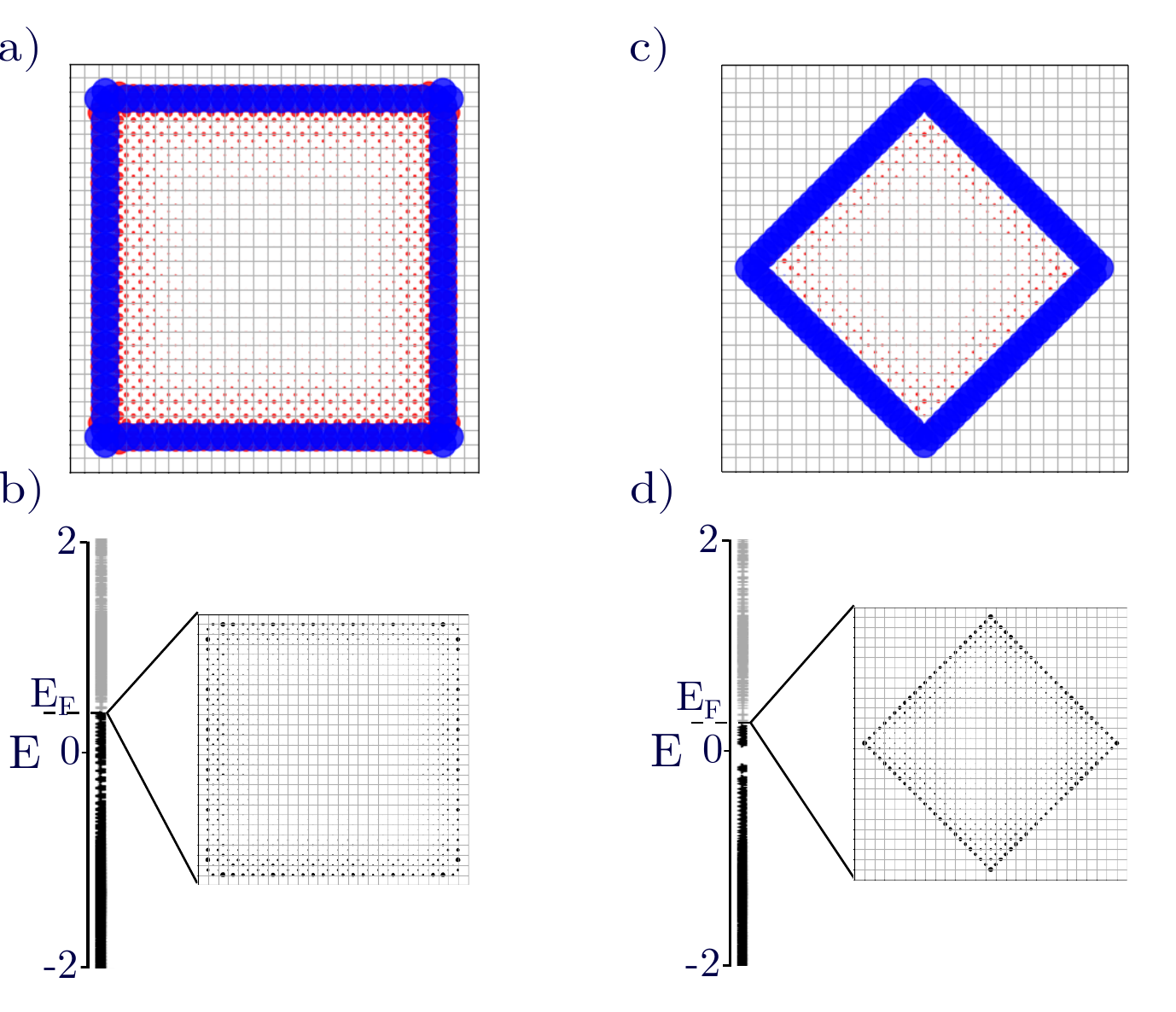}
    \caption{Corner charges for the straight cut (panels a and b) and diagonal cut (panels c and d) respectively for MSG75.5/the $k_z = 0$ plane of MSG77.18. On the top, we show the charge distribution, with Fermi level fixed to the same value as for MSG75.5, shown in Fig. \ref{fig:75_5_edge_charge}. Red indicates excess charge (relative to the center), blue a deficit of charge, where we sum over all occupied bands. On the bottom we show the spectrum, together with a state just below the Fermi level.  All calculations were done using the \texttt{PythTB} package \cite{PythTB2016}.}
    \label{fig:ap_75_5_edge_charge_spin_removed}
\end{figure}

We show the effect of removing a single spin on the boundary in Fig.~\ref{fig:ap_75_5_edge_charge_spin_removed}.

\section{Real-space invariants and twisted boundary conditions for MSG75.5}\label{ap:RSI/TBC_updated}
In this appendix, we provide further detail on the RSIs and TBCs discussed in section \ref{sec:RSI_TBC_main}. Our notation follows that of \cite{Twisted_BBC_theory} closely, and we use their tables throughout.

\subsection{Real space invariants}
Real-space invariants are defined at every WP as quantities which do not change as we move orbitals in a symmetry-preserving fashion from a high-symmetry WP to a lower-symmetry WP or vice versa. In the finite model of the $k_z = 0$ plane of MSG75.5, where the symmetries are broken down to wallpaper group p4, the relevant (non-magnetic) WPs are $1a$, $1b$ and $2c$, as shown in Fig.~\ref{fig:75_5_setup}a). Removing an orbital from WP $1a$ or $1b$ (with site-symmetry $C_4$) requires a minimum of four orbitals to come together, as this is the only way to consistently subduce to the trivial position. Thus, the imbalance between the four site-symmetry orbitals at WP $1a$ or $1b$ is protected, but not the total number of orbitals. This allows for the definition of three independent invariants (the real-space invariants) for each of these WP. For (non-magnetic) WP $2c$, with site-symmetry group $C_2$, pairs of orbitals must come together, so there is a single RSI. A full enumeration of the RSIs for wallpaper groups can be found in \cite{Twisted_BBC_theory}. For our model, we find the RSIs for the lower/upper subspace as: 
\begin{eqnarray}
\delta_{1a} &=& \pm 1  \nonumber\\
\delta_{2a} &=& \pm1  \nonumber\\
\delta_{3a} &=& \pm 1  \nonumber\\
\delta_{1b} &=& \mp 1  \nonumber\\
\delta_{2b} &=&0  \nonumber\\
\delta_{3b} &=& 0  \nonumber\\
\delta_{1c} &=& 0  \nonumber
\end{eqnarray}
 where $\delta_{iw}$ denotes RSI $i$ for the WP with label $w$. These RSIs can be used to directly confirm the fragility of our occupied manifold. In particular, for our symmetry setting (spinful $C_4$ without TRS), the criterion for fragility is that it be impossible to find numbers $N_{a}, N_{b}, N_{c}$ summing to $N_{\mathrm{bands}}=2$ whilst satisfying the constraints tabulated in \cite{Twisted_BBC_theory} and reproduced here for convenience:
\begin{eqnarray}\label{eq:orbital_constraints_P4_twisted}
N_{a/b} & = & \delta_{1a/1b}+\delta_{2a/2b}+\delta_{3a/3b} \ \mathrm{mod}\ 4  \nonumber \\
N_{a/b} & \geq & -3\delta_{1a/1b}+\delta_{2a/2b}+\delta_{3a/3b}  \nonumber\\
N_{a/b} & \geq & \delta_{1a/1b}-3\delta_{2a/2b}+\delta_{3a/3b}  \nonumber\\
N_{a/b} & \geq & \delta_{1a/1b}+\delta_{2a/2b}-3\delta_{3a/3b}  \nonumber\\
N_{a/b} & \geq & \delta_{1a/1b}+\delta_{2a/2b}+\delta_{3a/3b}  \nonumber\\
N_{c} & = & \delta_{1c} \ \mathrm{mod}\ 2  \nonumber\\
N_{c} & \geq & -\delta_{1c} \nonumber\\
N_{c} & \geq & \delta_{1c} \nonumber
\end{eqnarray}
In the occupied subspace, $N_{a}$ and $N_{b}$ have to equal $3 \ \mathrm{mod \ 4}$, whereas $N_{c}$ has to equal $0\ \mathrm{mod\ 2}$. However, this is not possible to satisfy if $N_a+N_b+N_c = 2$ Thus, the occupied subspace is indeed fragile. The RSIs in the unoccupied subspace, on the other hand, require $N_{a}$ and $N_{b}$ to equal $1\ \mathrm{mod\ 4}$, whilst $N_{c}$ still equals $0$ mod 2. This can be satisfied with $N_{a} = N_{b} = 1$ and $N_{c} = 0$. As can be easily checked, this also satisfies all other conditions implying that the unoccupied subspace is not fragile as expected.

\subsection{Twisted boundary conditions}
The twisted boundary conditions (TBC) are designed as perturbations to the Hamiltonian which leave the RSIs invariant but exchange the $C_4$ eigenvalues of bands \cite{Twisted_BBC_theory}. As the RSIs are defined in terms of the $C_4$ eigenvalues, if the relative balance of states with differing $C_4$ eigenvalues changes between the occupied and the unoccupied subspace, then there is necessarily a gap closing under this perturbation to ensure that the RSIs are invariant.

More concretely, as our system has $C_4$ symmetry, we can divide it into four regions which are related by symmetry, shown in Fig.~\ref{fig:TBC_combined}a). As we choose the cut illustrated in Fig.~\ref{fig:75_5_setup}c), there are no sites on the boundary between regions so each site is uniquely assigned to a region. Multiplying the hopping between the regions by some well-chosen parameter $\lambda$ (as described \cite{Twisted_BBC_theory}) amounts to a gauge change in the $C_4$ symmetry operator, permuting the site-symmetry representation. These are the TBC for this system. Using the RSIs, which are invariant under such transformations, we can predict which orbitals are exchanged. If chosen correctly, we can force an exchange between the occupied and the unoccupied spaces. The result of implementing the TBC by tuning $\lambda: 1\rightarrow i$ is shown in Fig.~\ref{fig:TBC_combined}. Note that the corner states (localized in the gap) are not cut by the boundary conditions and therefore do not flow. Checking the IRREPs of the states that exchange under TBC confirms that this pattern corresponds to our expected fragile phases, as explained in \cite{Twisted_BBC_theory}. This is thus another physical signature of the fragile magnetic topology in MSG75.5, which in this case is equivalent to fragility in the p4 wallpaper group. A similar pattern holds on the $k_z = 0$ plane of MSG77.18.

\end{document}